\documentclass[11pt,a4paper]{article}
\usepackage{jheppub}

\pdfoutput=1

\usepackage{tabularx}
\usepackage{amsmath,amssymb}
\usepackage{appendix}
\usepackage{graphicx}
\usepackage{color}
\usepackage{xcolor}
\usepackage{cancel}
\usepackage{subfigure}
\usepackage{multirow}
\usepackage{amssymb,changes}

\newcommand{\Rmnum}[1]{\expandafter\@slowromancap\romannumeral #1@}
\makeatother

\title{Gauged Q-ball dark matter through a cosmological first-order phase transition}
\author[a]{Siyu Jiang}
\affiliation[a]{MOE Key Laboratory of TianQin Mission, TianQin Research Center for
Gravitational Physics \& School of Physics and Astronomy, Frontiers
Science Center for TianQin, Gravitational Wave Research Center of CNSA, 
Sun Yat-sen University (Zhuhai Campus), Zhuhai 519082, China}
\emailAdd{jiangsy36@mail2.sysu.edu.cn}

\author[a]{Fa Peng Huang\footnote{Corresponding author.}}
\emailAdd{huangfp8@mail.sysu.edu.cn}

\author[b]{Pyungwon Ko}
\affiliation[b]{School of Physics, Korea Institute for Advanced Study, Seoul 02455, Korea}
\emailAdd{pko@kias.re.kr}

\abstract{As a new type of dynamical dark matter mechanism, we discuss the stability of the gauged Q-ball dark matter and its production mechanism through a cosmological first-order phase transition. This work delves into the study of gauged Q-ball dark matter generated during the cosmic phase transition. We demonstrate detailed discussions on the stability of gauged Q-balls to rigorously constrain their charge and mass ranges. Additionally, employing analytic approximations and the mapping method, we provide qualitative insights into gauged Q-balls. We establish an upper limit on the gauge coupling constant and give the relic density of stable gauged Q-ball dark matter formed during a first-order phase transition. Furthermore, we discuss potential observational signatures or constraints of gauged Q-ball dark matter, including astronomical observations and gravitational wave signals.}

\begin{document}
\maketitle

\section{Introduction}
Exploring the nature of dark matter (DM) is one of the central issues in (astro)particle physics and cosmology~\cite{Bertone:2016nfn}. So far, there are no expected signals of conventional DM candidates like Weakly Interacting Massive Particles (WIMPs) in the DM direct detection and collider search experiments~\cite{Boveia:2022adi,Boveia:2022syt,Cooley:2022ufh}. Then simple WIMP scenarios are strongly disfavored. 
There are a number of ways to save WIMP scenarios. For example, DM may have substantial couplings only to the 3rd generation fermions \cite{Baek:2016lnv,Baek:2017ykw,Abe:2016wck}, or dark sectors may consist 
of two or more stable DM species (see, for example, \cite{Khan:2023uii}). 
Or one can discard WIMP scenarios and consider other possibilities for DM productions and annihilations or decays in the early Universe. 
This status motivates us to study ultralight or ultra heavy DM candidate (for reviews, see Refs.~\cite{Baer:2014eja,Lin:2019uvt}).

Solitons produced in the early Universe are natural candidates of heavy DM 
(see, for example, Ref.~\cite{Baek:2013dwa} for hidden sector monopole DM accompanied by stable spin-1 vector DM and massless dark radiation, and Ref.~\cite{Derevianko:2013oaa} for hunting for topological DM using atomic clocks).
These solitons are specific field configurations which are classified into two classes, namely, the topological solitons and the non-topological solitons. Recently, as renaissance of the quark nuggets DM proposed by Witten~\cite{Witten:1984rs}, various new ideas on the non-topological soliton DM are proposed, where the DM relic density can be produced by the dynamical process of cosmological first-order phase transition (FOPT), such as the Q-ball DM~\cite{Krylov:2013qe,Huang:2017kzu,Jiang:2023qbm,Hong:2020est}. These new mechanisms can naturally avoid the unitarity problem for heavy DM~\cite{Griest:1989wd}.
Dynamical DM mechanisms are specified by the DM penetration behavior into the bubble which depends on the DM mass and bubble wall velocity~\cite{Baker:2019ndr,Chway:2019kft,Azatov:2021ifm}. Phase transitions in the early Universe can also be the source of primordial black holes~\cite{Khlopov:2000js,Dymnikova:2000dy,Kanemura:2024pae}.

There are extensive discussions on the non-topological solitons in a theory of complex scalar field with global $U(1)$ symmetry, proposed in~\cite{Rosen:1968mfz} and known as Q-balls~\cite{Coleman:1985ki}. And it is natural to study the Q-balls in the gauged case~\cite{Coleman:1985ki,Lee:1988ag,Rosen:1968zwl,Lee:1991bn,Friedberg:1976me,Arodz:2008nm,Benci:2010cs,Benci:2012fra,Dzhunushaliev:2012zb}, 
by promoting the global $U(1)$ symmetry to the 
local gauge $U(1)$ symmetry. For reviews of $U(1)$ gauged Q-balls, 
see \cite{Gulamov:2013cra,Gulamov:2015fya,Nugaev:2019vru}. 
Q-balls have been proposed as a potential DM candidate in supersymmetric theories~\cite{Kusenko:1997zq,Kusenko:1997si}. They can also explain the baryon asymmetry of the Universe~\cite{Kasuya:2012mh}.
The gauged Q-ball DM in supersymmetry model has 
been studied in several papers~\cite{Hong:2017qvx,Hong:2016ict}.
It is meaningful to search for other production mechanisms of Q-ball or gauged Q-ball DM without supersymmetry.
In this paper, we study whether the $U(1)$
gauged Q-balls produced during cosmic phase transition could be a viable DM candidate. If the gauged Q-balls can be stable under certain
circumstances, we still need some mechanism to
(1) produce the charge asymmetry (i.e. locally produce lots of particles with the same charge to form Q-ball)
(2) and packet the same sign charge in the small size after overcoming the Coulomb repulsive interaction. For the first condition, the primordial 
charge asymmetry could be produced by some early Universe processes such as decays  
of heavier particles. Cosmological FOPT can naturally realize the second condition and can produce phase transition gravitational wave (GW) which can be detected by future GW experiments, such as LISA~\cite{LISA:2017pwj},
TianQin~\cite{TianQin:2015yph,Liang:2022ufy}, Taiji~\cite{Hu:2017mde}, BBO~\cite{Corbin:2005ny}, DECIGO~\cite{Seto:2001qf}, and Ultimate-DECIGO~\cite{Kudoh:2005as}.

In this work, for the first time, we study the natural production mechanism of gauged Q-ball DM through a cosmological FOPT. The paper is organised as follows. We describe the basic model that can produce the gauged Q-balls and the numerical solutions of the Q-ball profiles in section~\ref{gaugedqball}. Basic properties and the stable parameter space of gauged Q-balls are discussed in section~\ref{qballproperty}. Thin-wall approximation and the corresponding analytic evaluations are given in section~\ref{thinwall}. Phase transition dynamics in the Standard Model (SM) plus an extra singlet and the relic density of gauged Q-ball DM are elucidated in section~\ref{FOPTDM}. Signals and constraints of gauged Q-ball DM are given in section~\ref{signalconstraints}. Concise conclusions and discussions are given in section~\ref{conclusion}.


\section{Gauged Q-ball}\label{gaugedqball}
\subsection{Friedberg-Lee-Sirlin--Maxwell model}
In this work, we adopt the Friedberg-Lee-Sirlin two-component model~\cite{Friedberg:1976me}\footnote{The Friedberg-Lee-Sirlin two-component model has been reviewed in detail in Refs.~\cite{Heeck:2023idx,Ponton:2019hux}.} plus gauge component, which is called Friedberg-Lee-Sirlin-Maxwell (FLSM) model~\cite{Lee:1991bn}. This model and the corresponding stability of $U(1)$ gauged Q-ball have been discussed in~\cite{Lee:1988ag,Lee:1991bn,Loiko:2019gwk,Loiko:2022noq,Kinach:2022jdx}.
We begin our discussions with the following Lagrangian density
\begin{equation}\label{zerolag}
\mathcal{L}_{}=\left(D_\mu \phi\right)^{\dagger}\left(D^\mu \phi\right)+\frac{1}{2}\partial_\mu h \partial^\mu h-\frac{1}{4}  \tilde A_{\mu \nu} \tilde A^{\mu \nu}-V(\phi,h)\,\,,
\end{equation}
where the potential $V(\phi,h)$ reads
\begin{equation}
V(\phi,h) = \frac{\lambda_{\phi h}}{2} h^2|\phi|^2+\frac{\lambda_h}{4}\left(h^2-v_0^2\right)^2\,\,.
\end{equation}
$\phi$ and $h$ are the complex scalar field and (real) Higgs field respectively.
$D_\mu =\partial_\mu+i \tilde{g} \tilde A_\mu$ and $\tilde A_{\mu \nu}=\partial_\mu \tilde A_\nu-\partial_\nu \tilde A_\mu$ where $\tilde A_\mu$ is a dark $U(1)$ gauge field and $\tilde{g}$ is the corresponding gauge coupling constant. $\tilde A_\mu$ can be identified as the dark electromagnetic field. We fix the Higgs mass $m_h = 125~\mathrm{GeV}$ and vacuum expectation value $v_0 = 246 ~\mathrm{GeV}$ at zero temperature then $\lambda_h = m_h^2/(2v_0^2) \approx 0.13$. 
The complex scalar $\phi$ gains mass through the portal coupling with the Higgs. 
In the true vacuum, $m_\phi=\sqrt{\frac{\lambda_{\phi h}}{2}} v_0$.
We assume $\lambda_{\phi h}>0$, and thus
the Lagrangian density is symmetric under the dark $U(1)$ symmetry which remains unbroken when the Universe goes through the electroweak phase transition.
The local $U(1)$ gauge symmetry leads to the conserved current, 
\begin{equation}
J_\mu=i\left(\phi^{\dagger} \overleftrightarrow{\partial}_\mu \phi+2 i \tilde{g} \tilde A_\mu|\phi|^2\right)\,\,,
\end{equation}
and the corresponding conserved charge,
\begin{equation}
Q=\int d^3 x J^0\,\,.
\end{equation}

Once the gauged Q-balls are formed in this FLSM model, one could consider a coherent configuration of $\phi$, $h$, and $\tilde A_\mu$ at a given charge $Q$. 
The lowest energy state will have no ``magnetic field" so the space component $\tilde A_i=0$~\cite{Lee:1988ag,Lee:1991bn}. We assume spherical symmetry for the lowest energy configuration.  Scaling away the physical dimensions, we introduce dimensionless field variables $\mathcal{A}, \Phi$, and $\mathcal{H}$ defined in the convention of Ref.~\cite{Lee:1991bn}.
\begin{equation}\label{convention}
\tilde A_{t}(r)=v_0 \frac{\tilde{g}}{\sqrt{2\lambda_h}} \mathcal{\mathcal{A}}(\rho), \quad \phi(t,r)=\frac{v_0}{\sqrt{2}} \Phi(\rho) e^{-i \omega t}, \quad h(r)=v_0 \mathcal{H}(\rho)\,\,,
\end{equation}
where $\rho \equiv \sqrt{2\lambda_h}  v_0 r= m_h r$. The Lagrangian, with the substitution of the field variables defined above, becomes
\begin{equation}\label{lag}
\begin{aligned}
	L= & -4 \pi \frac{v_0}{\sqrt{2\lambda_h}} \int d \rho \rho^2\left[\frac{1}{2}\left(\nu-\alpha^2 \mathcal{A}\right)^2 \Phi^2-\frac{1}{8}\left(\mathcal{H}^2-1\right)^2-\frac{k^2}{2} \mathcal{H}^2 \Phi^2-\frac{1}{2}\left(\partial_\rho \Phi\right)^2\right. \\
	& \left.-\frac{1}{2}\left(\partial_\rho \mathcal{H}\right)^2+\frac{\alpha^2}{2}\left(\partial_\rho \mathcal{A}\right)^2\right],
\end{aligned}
\end{equation}
where $\alpha \equiv \frac{|\tilde{g}|}{\sqrt{2\lambda_h} }, k \equiv \frac{\sqrt{\lambda_{\phi h}}}{2\sqrt{\lambda_h} }=\frac{m_\phi}{m_h}$, and $\nu \equiv \frac{\omega}{\sqrt{2\lambda_h}  v_0}$.
By varying $L$ with respect to $\mathcal{A}, \Phi$, and $\mathcal{H}$, we find the equations of motion (EoM) for the three fields,
\begin{equation}\label{Ap}
		\frac{1}{\rho^2} \partial_\rho\left(\rho^2 \partial_\rho \mathcal{A}\right)+ (\nu-\alpha^2 \mathcal{A}) \Phi^2=0\,\,,
\end{equation}
\begin{equation}\label{Bp}
	\frac{1}{\rho^2} \partial_\rho\left(\rho^2 \partial_\rho \Phi\right)+\left[(\nu-\alpha^2 \mathcal{A})^2-k^2 \mathcal{H}^2\right] \Phi=0\,\,,
\end{equation}
and
\begin{equation}\label{Cp}
	\frac{1}{\rho^2} \partial_\rho\left(\rho^2 \partial_\rho \mathcal{H}\right)-k^2 \mathcal{H} \Phi^2-\frac{1}{2} \mathcal{H}\left(\mathcal{H}^2-1\right)=0\,\,.
\end{equation}

The total energy is given by
\begin{equation}\label{energy}
E=\frac{4 \pi v_0}{\sqrt{2\lambda_h}} \int d \rho \rho^2 \mathcal{E}\,\,,
\end{equation}
where $\mathcal{E}=\frac{\alpha^2}{2 }\left(\partial_\rho \mathcal{A}\right)^2+\frac{1}{2}\left(\partial_\rho \Phi\right)^2+\frac{1}{2}\left(\partial_\rho \mathcal{H}\right)^2+\frac{1}{2}\left[(\nu-\alpha^2 \mathcal{A})^2+k^2 \mathcal{H}^2\right] \Phi^2+\frac{1}{8}\left(\mathcal{H}^2-1\right)^2$. 
And the total charge is given by
\begin{equation}\label{chag}
Q=\frac{2 \pi}{\lambda_h} \int d \rho \rho^2 (\nu-\alpha^2 \mathcal{A}) \Phi^2\,\,.
\end{equation}
From Eqs.~\eqref{Ap} and \eqref{chag} we see that
\begin{equation}
\frac{\lambda_h Q}{2\pi}=-\lim _{\rho \rightarrow \infty} 4 \pi \rho^2 \partial_\rho \mathcal{A}\,\,.
\end{equation}
Therefore we get, for large $\rho$, $\mathcal{A} \rightarrow \frac{\lambda_h Q}{2 \pi  \rho}$ or $\tilde A_t\rightarrow \frac{\tilde{g}Q}{4\pi r}$. Then Eq.~\eqref{Bp} at large $\rho$ becomes  
\begin{equation}\label{largeB}
\frac{1}{\rho^2} \partial_\rho\left(\rho^2 \partial_\rho \Phi\right)-\frac{2\nu \tilde{g}^2Q}{4\pi \rho}\Phi+\left(\nu^2-k^2\right) \Phi=0\,\,.
\end{equation}
It has been shown in Ref.~\cite{Gulamov:2015fya} for $\nu<k$  this equation at $\rho \rightarrow \infty$ has the solution of the form,
\begin{equation}
\Phi(\rho)=C_U \mathrm{e}^{-\sqrt{k^2-\nu^2 }\rho}~ U\left(1+\frac{\nu \tilde{g}^2 Q}{4 \pi \sqrt{k^2-\nu^2}}, 2,2 \sqrt{k^2-\nu^2} \rho\right)\,\,,
\end{equation}
where $C_U$ is a constant and $U(a, b, z)$  is the confluent hypergeometric 
function of the second kind. For $\sqrt{k^2-\nu^2} \rho \gg 1$ we get
\begin{equation}
\Phi(\rho) \sim \rho^{-1-\frac{\nu \tilde{g}^2 Q}{4 \pi \sqrt{k^2-\nu^2}}}\mathrm{e}^{-\sqrt{k^2-\nu^2}\rho}\,\,.
\end{equation}
We can see that in the limit $\tilde{g}\rightarrow 0$ this form coincides with the nongauged global Q-ball, $\Phi(\rho) \sim \rho^{-1}\mathrm{e}^{-\sqrt{k^2-\nu^2}\rho} $. The difference is caused by taking into account the dark electromagnetic potential $\mathcal{A}(\rho)$. For $\omega=m_\phi$ or $\nu = k$, Eq. \eqref{largeB} takes the form of
\begin{equation}
	\frac{1}{\rho^2} \partial_\rho\left(\rho^2 \partial_\rho \Phi\right)-\frac{2k \tilde{g}^2Q}{4\pi \rho}\Phi=0\,\,.
\end{equation}
The solution to this equation reads
$
\Phi(\rho)=C_K \frac{K_1\left(\sqrt{\frac{2 k \tilde{g}^2 Q}{\pi} \rho}\right)}{\sqrt{\rho}},
$
where $C_K$ is a constant and $K_1(z)$ is the modified Bessel function of the second kind. For large $\rho$ this solution has the form of
$
\Phi(\rho) \sim {\rho^{-\frac{3}{4}}}\mathrm{e}^{-\sqrt{\frac{2 k \tilde{g}^2 Q}{\pi} \rho}}
$
which also differs from the global case, in which one expects $\Phi(\rho) \sim \frac{1}{\rho}$ for $\nu=k$. For the case $\nu>k$, it can be seen from Eq. \eqref{largeB} that the corresponding solutions for the complex scalar field are oscillatory at $\rho \rightarrow \infty$, leading to unexpected infinite charge and energy. These solutions are unphysical and should be discarded.

It is convenient to write Eq.~\eqref{Ap} in the form of
\begin{equation}\label{partialA}
\partial_\rho\left(\rho^2 \partial_\rho \mathcal{A}\right)=- (\nu-\alpha^2 \mathcal{A}) \Phi^2\rho^2\,\,.
\end{equation}
Suppose that $\mathcal{A}(0)>\nu/\alpha^2$. Eq.~\eqref{partialA} then implies that $\rho^2 \partial_\rho \mathcal{A}$ is an increasing function of $\rho$ such that $\partial_\rho \mathcal{A}>0$ and $\mathcal{A}(\rho)>\nu/\alpha^2$ for all $\rho>0$. This possibility is not acceptable, given that $\nu>0$ and $\mathcal{A}(\rho) \rightarrow 0$ at $\rho \rightarrow \infty$. The only acceptable possibility is that $\mathcal{A}(0)\leq\nu/\alpha^2$. Then $\partial_\rho \mathcal{A}<0$ and therefore $\mathcal{A}(\rho)$ is a monotonically decreasing function of $\rho$. We can then say that $\mathcal{A}(\rho)$ obeys the inequalities
\begin{equation}
0 \leq\alpha^2 \mathcal{A}(\infty) \leq \alpha^2 \mathcal{A}(\rho) \leq\alpha^2 \mathcal{A}(0)\leq \nu\leq  k\,\,.
\end{equation}

The energy integral Eq.~\eqref{energy} can be written in a different form. Demanding that the Lagrangian \eqref{lag} is stationary at $\epsilon=1$ under the rescaling of the form $\rho \rightarrow \epsilon \rho$,  leading to the relation $dL(\rho \rightarrow \epsilon \rho)/d \epsilon|_{\epsilon=1}=0$, this gives the relation:
\begin{equation}\label{rescal}
\begin{aligned}
	& 3 \int d \rho \rho^2\left[\frac{1}{2} (\nu-\alpha^2 \mathcal{A})^2 \Phi^2-\frac{1}{8}\left(\mathcal{H}^2-1\right)^2-\frac{1}{2} \mathcal{H}^2 \Phi^2 k^2\right] \\
	= & \int d \rho \rho^2\left[\frac{1}{2}\left(\partial_\rho \Phi\right)^2+\frac{1}{2}\left(\partial_\rho \mathcal{H}\right)^2-\frac{\alpha^2}{2 }\left(\partial_\rho \mathcal{A}\right)^2\right]\,\,.
\end{aligned}
\end{equation}

Substituting Eqs.~\eqref{rescal} and \eqref{Ap} into Eq.~\eqref{energy}, one obtains
\begin{align}\label{engQ}
E&= \frac{4 \pi v_0}{\sqrt{2\lambda_h}} \int d \rho\left[(\nu-\alpha^2 \mathcal{A})^2 \Phi^2+\frac{1}{3}\left(\partial_\rho \Phi\right)^2+\frac{1}{3}\left(\partial_\rho \mathcal{H}\right)^2+\frac{2\alpha^2}{3 }\left(\partial_\rho \mathcal{A}\right)^2\right]\notag \\
 &=\frac{4 \pi v_0}{\sqrt{2\lambda_h}} \int d \rho\left[-(\nu-\alpha^2 \mathcal{A}) \partial_\rho\left(\rho^2 \partial_\rho \mathcal{A}\right)+\rho^2\left\{\frac{1}{3}\left(\partial_\rho \Phi\right)^2+\frac{1}{3}\left(\partial_\rho \mathcal{H}\right)^2+\frac{2\alpha^2}{3 }\left(\partial_\rho \mathcal{A}\right)^2\right\}\right] \notag \\
 &=\omega Q+\frac{4 \pi}{3} \frac{v_0}{\sqrt{2\lambda_h}} \int d \rho \rho^2\left[\left(\partial_\rho \Phi\right)^2+\left(\partial_\rho \mathcal{H}\right)^2-{\alpha^2}\left(\partial_\rho \mathcal{A}\right)^2\right]\,\,,
\end{align}
where in the third line we have integrated by part and use the fact that $\rho^2 \partial_\rho \mathcal{A} \rightarrow -\frac{\tilde{g}^2 Q}{4 \pi \alpha^2}$ and $\mathcal{A} \rightarrow 0 $ at large $\rho$. It can be seen that the energy
for the free field solution when we neglect the variation of $\Phi$ and $\mathcal{H}$ takes the form
\begin{equation}
E_{\text {free }}=m_\phi Q+\text { ``dark electrostatic energy" }\,\,.
\end{equation}
The ``dark electrostatic energy" is roughly proportional to $\frac{\tilde{g}^2 Q^2}{R}$ for charge uniformly distributed on scale $R$.

\subsection{Numerical results of the field configuration, energy, and charge}
\begin{figure*} [t!]
	\centering
	\subfigure{
		\includegraphics[scale=0.45]{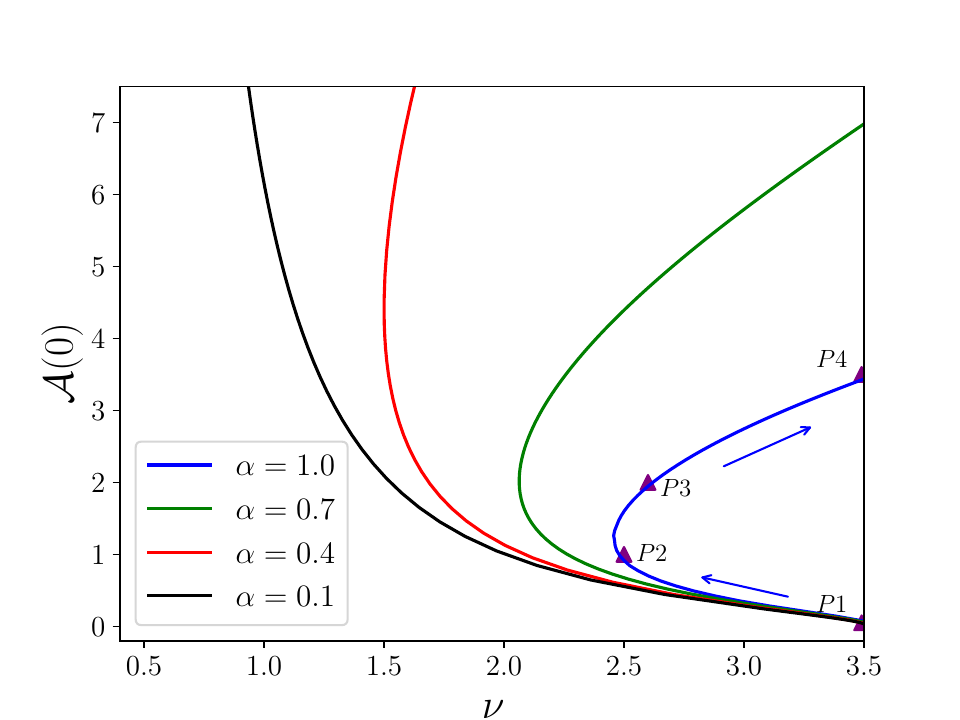}}
	\subfigure{
		\includegraphics[scale=0.45]{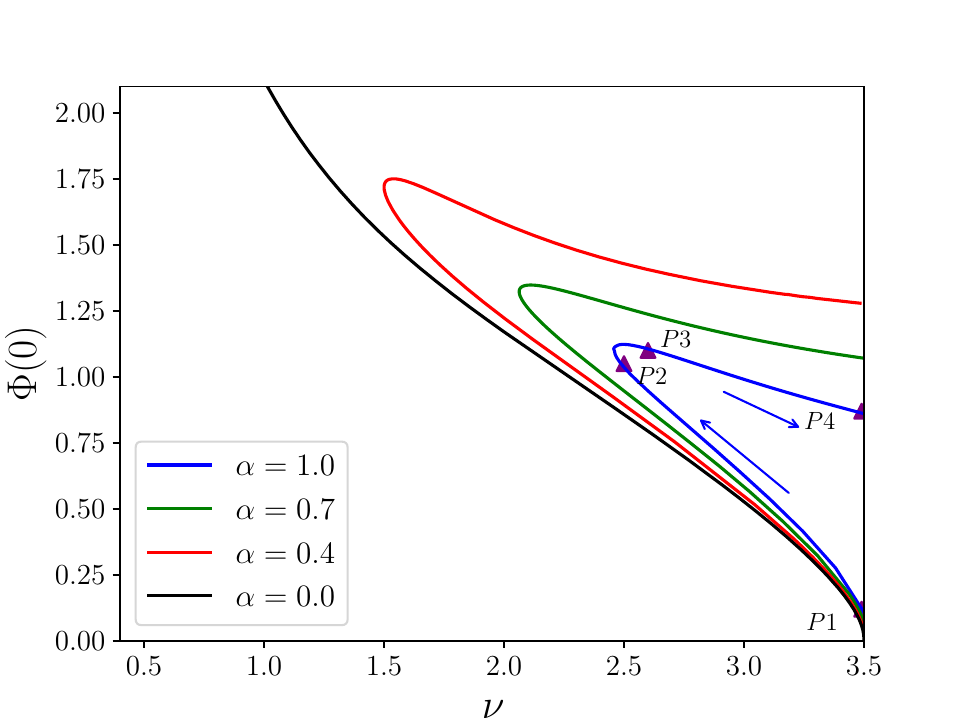}}
	\\
	\subfigure{
		\includegraphics[scale=0.45]{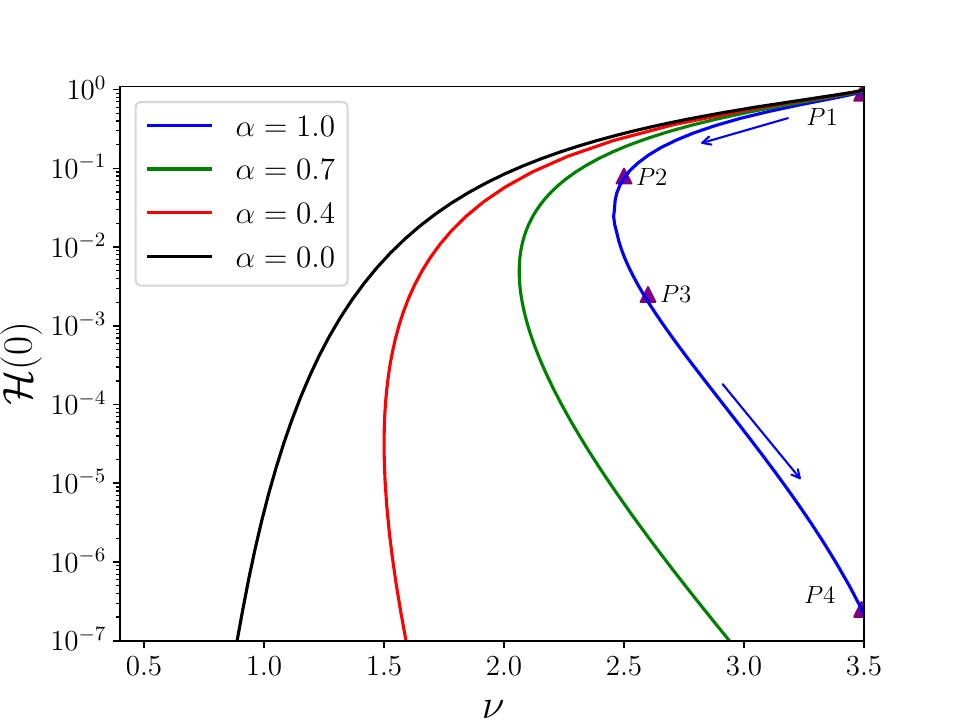}}
	\subfigure{
		\includegraphics[scale=0.45]{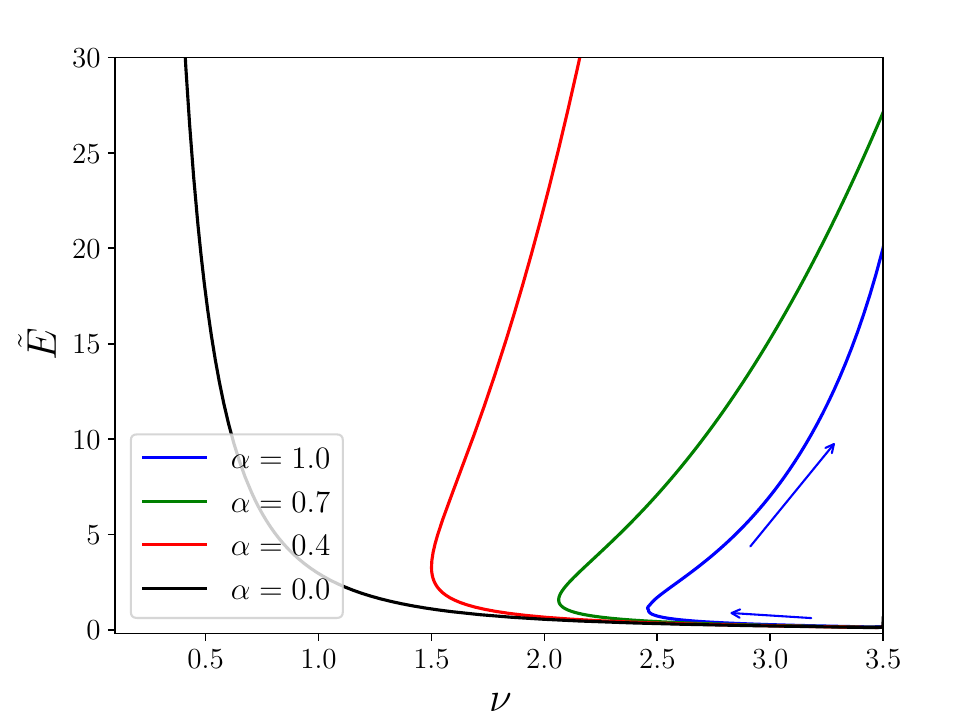}}
	\caption{Values of the three fields at the Q-ball center and the total energy $\tilde E$ for different values of gauge coupling $\alpha$. The arrows represent the evolution of the frequency $\nu$. We choose $k=3.5$ which corresponds to $\lambda_{\phi h}\approx 6$. Four specific solutions $P1, P2, P3, P4$ for $\alpha=1.0$ are marked by the triangles and the corresponding profiles are shown in figure \ref{fieldpro}.}\label{field0}
\end{figure*}
After qualitative analysis of the gauged Q-ball solution, we begin to numerically solve Eqs.~\eqref{Ap}, \eqref{Bp}, and \eqref{Cp} with the following boundary conditions,
\begin{equation}
\frac{\partial \mathcal{A}}{\partial \rho}=\frac{\partial \Phi}{\partial \rho}=\frac{\partial \mathcal{H}}{\partial \rho}=0 \text { at } \rho=0,\quad  \mathcal{A}=\Phi=0 \text { and } \mathcal{H}=1 \text { at } \rho=\infty\,\,.
\label{eq:bc} 
\end{equation}
The first boundary condition is necessary so that the terms $\frac{2}{\rho}\left(\frac{\partial \mathcal{A}}{\partial \rho}\right), \frac{2}{\rho}\left(\frac{\partial \Phi}{\partial \rho}\right)$, and $\frac{2}{\rho}\left(\frac{\partial \mathcal{H}}{\partial \rho}\right)$ do not become singular at $\rho=0$, and the latter is necessary because the energy density $\mathcal{E}$ and charge density $(\nu-\alpha^2 \mathcal{A}) \Phi^2$ should be integrable 
over the infinite 
spatial volume and the integral should be finite.

It is convenient to introduce the dimensionless quantities for the total energy and charge of the gauged Q-ball,  
\begin{equation}
	\begin{aligned}
		\tilde{E} & \equiv \frac{\lambda_h }{2 \pi m_\phi}E=\frac{1}{k} \int_0^{\infty} d \rho \rho^2 \mathcal{E}\,\,, \\
		\tilde{Q} & \equiv \frac{\lambda_h Q}{2 \pi}=\int_0^{\infty} d \rho \rho^2 (\nu-\alpha^2 \mathcal{A}) \Phi^2 \,\,,
	\end{aligned}
\end{equation}
which can be calculated directly once the numerical solutions of the corresponding differential equations are found. 

We use the relaxation method~\cite{NumericalRecipes,Guo:2020tla} to solve coupled 2nd-order ordinary differential equations, Eqs.~\eqref{Ap}, \eqref{Bp} and \eqref{Cp}, with boundary conditions, Eq.~\eqref{eq:bc}. The relaxation method solves the boundary value problems by updating the trial functions on the grid in an iterative way. As an example, we fix $k=3.5$ and we scan over all of the solutions at a given value of $\alpha$. The results can be seen in figure \ref{field0}. The frequency $\nu$ firstly decreases in the direction of the arrow. We call this the ``first branch" where the backreaction of gauge field is small. Then the solutions turn on the ``second branch" as $\nu$ increases and the gauge field dominates. Contrary to the global Q-balls, the parameter $\nu$ does not uniquely determine the charge and energy of the gauged Q-balls. For the global case, the energy and charge increase as the $\nu$ approaches zero. However, in the case of gauged Q-balls, the $\nu$ is replaced by $(\nu-\alpha^2\mathcal{A})$. Then on the second branch where the gauge field $\mathcal{A}$ dominates, $\nu$ has to increase in order to satisfy $(\nu-\alpha^2 \mathcal{A}) > 0$.

\begin{figure}[h]
	\begin{minipage}{0.326\linewidth}
		\vspace{3pt}
		\centerline{\includegraphics[width=\textwidth]{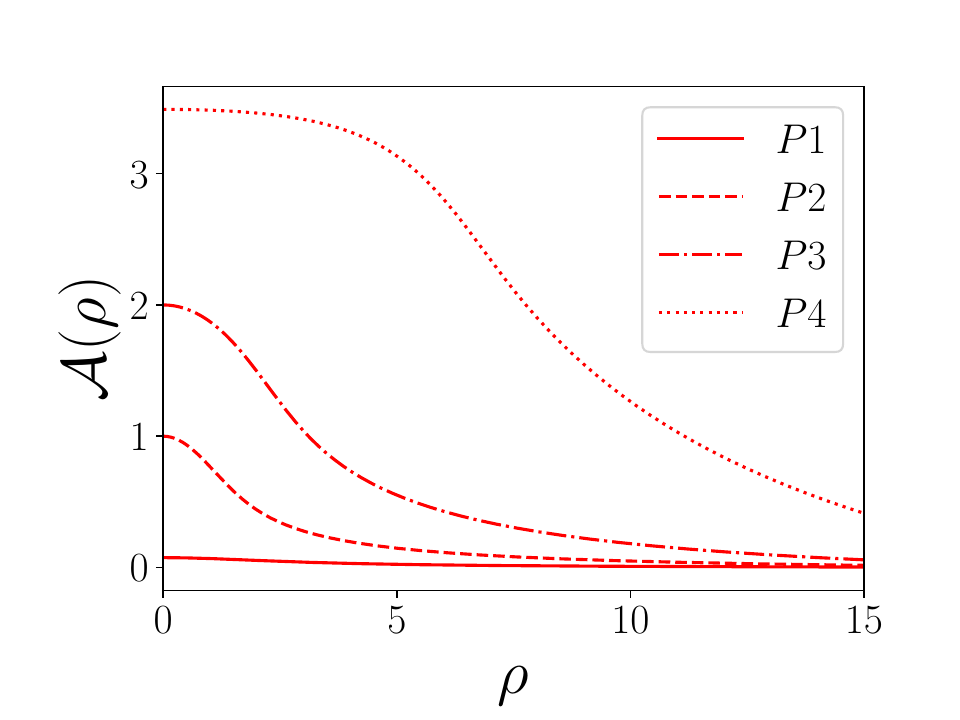}}
	\end{minipage}
	\begin{minipage}{0.326\linewidth}
		\vspace{3pt}
		\centerline{\includegraphics[width=\textwidth]{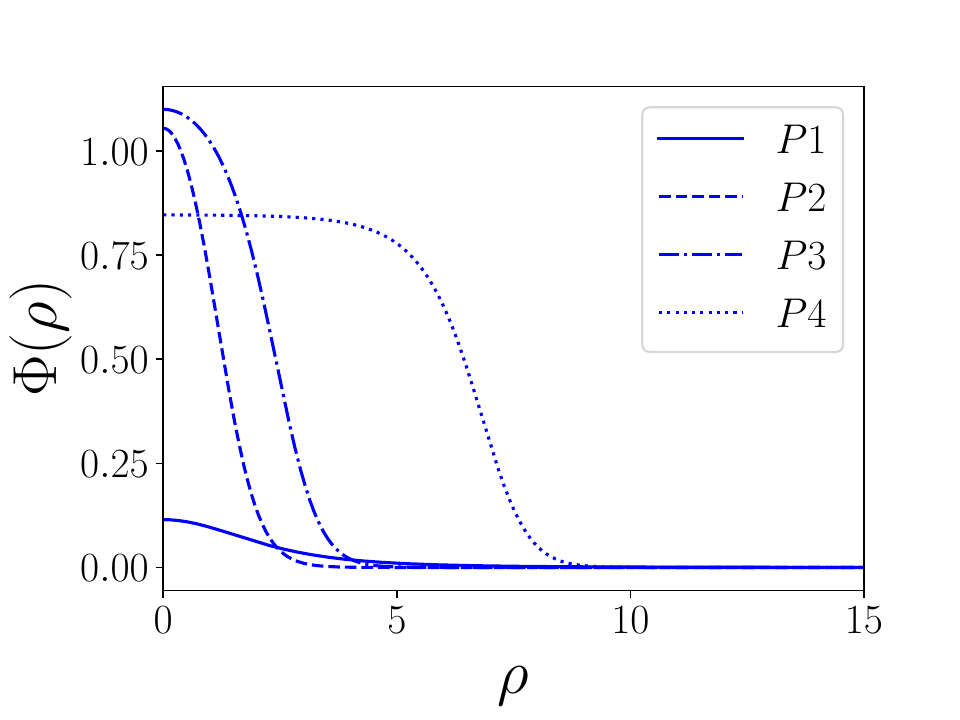}}
	\end{minipage}
	\begin{minipage}{0.326\linewidth}
		\vspace{3pt}
		\centerline{\includegraphics[width=\textwidth]{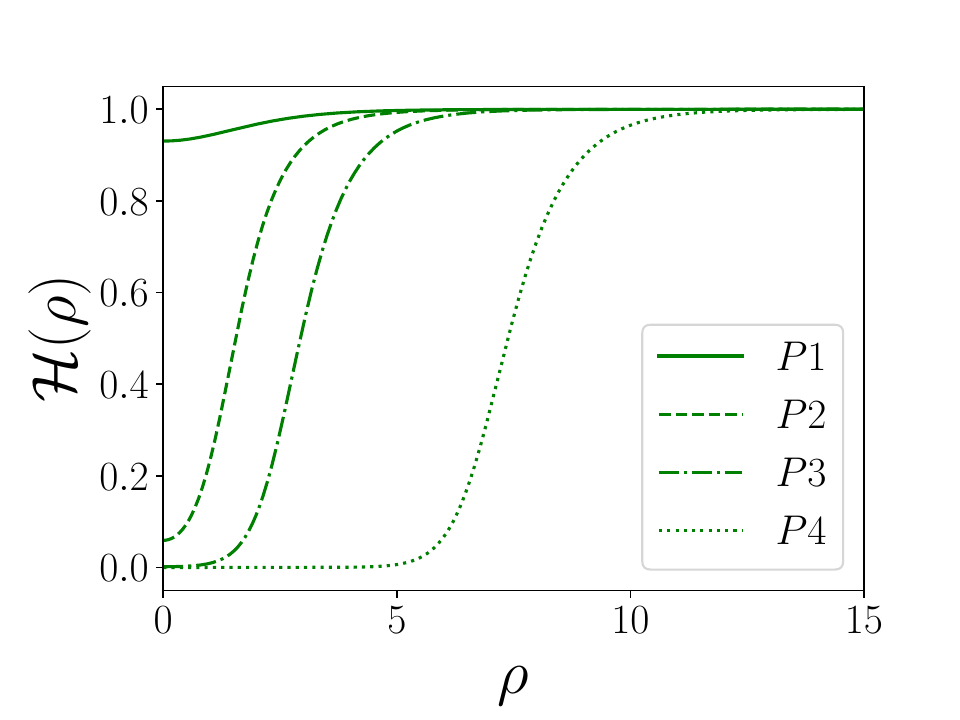}}
	\end{minipage}
	\caption{Profiles of the dark gauge field, complex scalar field, Higgs field of the gauged Q-ball. Here we choose the marked points $P1, P2, P3, P4$ in figure \ref{field0} where $k=3.5$ and $\alpha=1.0$.}
	\label{fieldpro}
\end{figure}

We choose four specific solutions $P1, P2, P3, P4$ in figure \ref{field0} for $\alpha=1.0$ which are marked by the purple triangles and the corresponding numerical profiles of $\mathcal{A}$, $\Phi$ and $\mathcal{H}$ are shown in figure \ref{fieldpro}. We can see that as the value of $\mathcal{A}(0)$ of the gauged Q-ball becomes larger, the Higgs field value inside the Q-ball is closer to zero. Actually, when the value of $\mathcal{A}(0)$ increases, the radius, charge and energy also increase. In other words, the Higgs value is effectively zero inside for large gauged Q-balls.

\begin{figure}[h]
	\begin{minipage}{0.48\linewidth}
		\vspace{3pt}
		\centerline{\includegraphics[width=\textwidth]{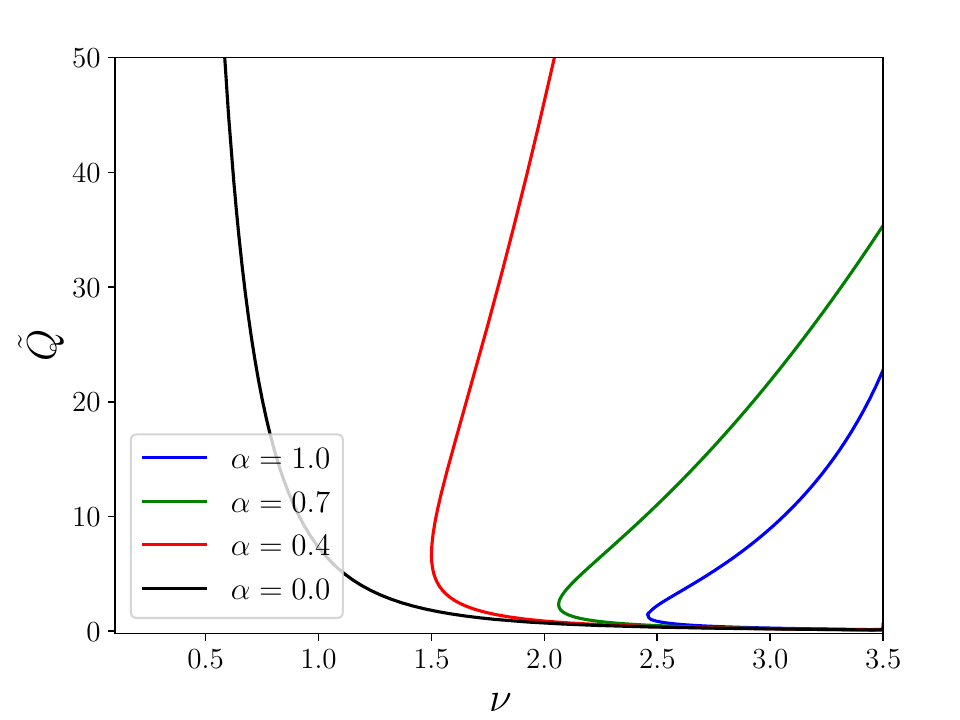}}
	\end{minipage}
	\begin{minipage}{0.52\linewidth}
		\vspace{3pt}
		\centerline{\includegraphics[width=\textwidth]{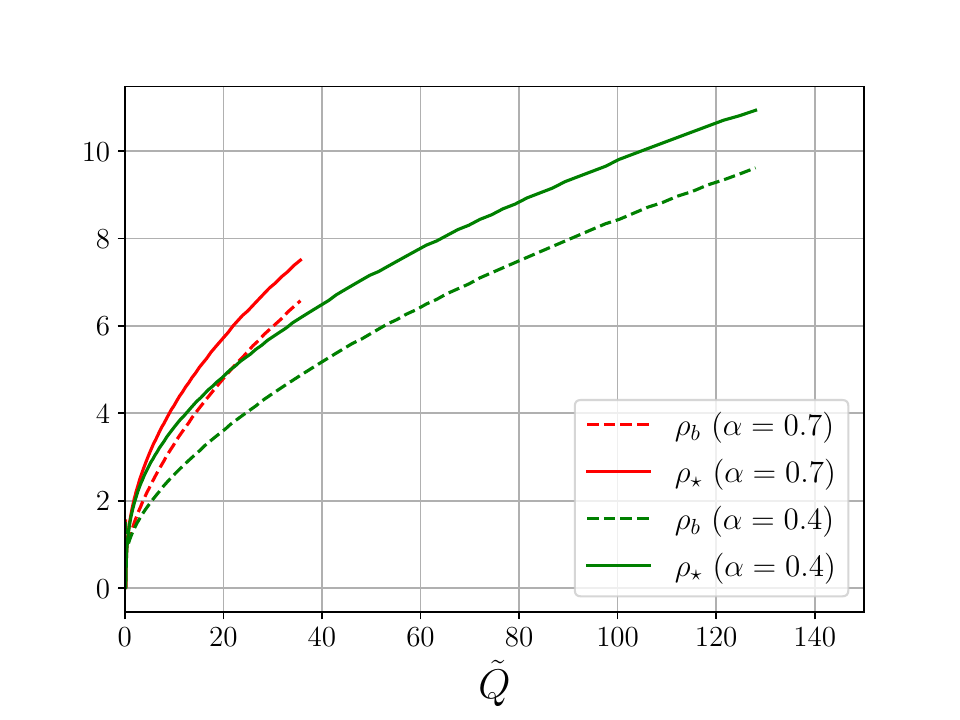}}
	\end{minipage}
	\caption{Left: total charge $\tilde Q$ for different values of $\alpha$ for $U(1)$ gauged Q-balls. Right: the two typical radii for the Q-ball field $\Phi(\rho)$ and Higgs field $\mathcal{H}(\rho)$.}
	\label{rsrb}
\end{figure}

The total charge of gauged Q-balls for different values of gauge coupling $\alpha$ is shown in the left panel of figure \ref{rsrb}. It can be seen that the charge of the gauged Q-balls is also finite at a nonzero $\alpha$ whereas for the global Q-balls the charge is unbounded from above. In order to obtain gauged Q-balls with relatively large charge, the gauge coupling has to be small enough.

We can define two typical radii for the Higgs field $\mathcal{H}(\rho)$ and the Q-ball field $\Phi(\rho)$ respectively. The $\rho_\star$ is defined by $\mathcal{H}(\rho_\star) =1/2$ and $\rho_b$ is defined by $\Phi(\rho_b) = \Phi(0)/2$. These two radii are shown in the right panel of figure \ref{rsrb}. We can see that $\rho_\star$ is generally larger than $\rho_b$. One may prefer to call $\rho_\star$ the radius of the ``Higgs ball". Hereafter, we use $\rho_\star$ to represent the Q-ball radius which defines the charge and energy of gauged Q-balls.

\section{Basic properties of gauged Q-balls}\label{qballproperty}

\subsection{$dE/dQ$ for gauged Q-balls}
It is well known that for non-gauged global Q-balls the relations $dE/dQ=\omega$ holds. Here we will show that this also holds for gauged Q-balls in FLSM model. From Eq.~\eqref{energy}, we have
\begin{align}
	\frac{ dE}{d\nu}
	 &= \frac{4 \pi v_0}{\sqrt{2\lambda_h}} \int d \rho \rho^2 \left\{\alpha^2\left(\partial_\rho \mathcal{A}\right)\left(\partial_\rho \frac{d \mathcal{A}}{d\nu}\right)+\left(\partial_\rho \Phi\right)\left(\partial_\rho \frac{d\Phi}{d\nu}\right)+\left(\partial_\rho \mathcal{H}\right)\left(\partial_\rho \frac{d \mathcal{H}}{d\nu}\right) \right.\notag\\
	 &+\left[\left(1-\alpha^2 \frac{d \mathcal{A}}{d\nu}\right)(\nu-\alpha^2 \mathcal{A})+k^2 \mathcal{H} \frac{d \mathcal{H}}{d\nu}\right] \Phi^2+\left[(\nu-\alpha^2 \mathcal{A})^2+k^2 \mathcal{H}^2 \right] \Phi\frac{d \Phi}{d\nu} \notag\\ &\left. +\frac{1}{2}\left(\mathcal{H}^2-1\right)\mathcal{H}\frac{d \mathcal{H}}{d\nu}\right\}\,\,.
\end{align}
After integrating $\partial_{\rho} \Phi$ and $\partial_{\rho} \mathcal{H}$ by part and using Eqs.~\eqref{Bp} and \eqref{Cp}, we can get
\begin{equation}
	\begin{aligned}
		\frac{ dE}{d\nu}
		&= \frac{4 \pi v_0}{\sqrt{2\lambda_h}} \int d \rho \rho^2 \left[\alpha^2\left(\partial_\rho \mathcal{A}\right)\left(\partial_\rho \frac{d \mathcal{A}}{d\nu}\right) + (\nu-\alpha^2 \mathcal{A})\left(1-\alpha^2 \frac{d \mathcal{A}}{d\nu}\right)\Phi^2 +2(\nu-\alpha^2 \mathcal{A})^2\Phi\frac{d\Phi}{d\nu} \right] \\
		&= \frac{4 \pi v_0}{\sqrt{2\lambda_h}} \int d \rho \rho^2 \left\{(\nu-\alpha^2 \mathcal{A})\frac{d}{d\nu}\left[(\nu-\alpha^2 \mathcal{A})\Phi^2\right]+\alpha^2\left(\partial_\rho \mathcal{A}\right)\left(\partial_\rho \frac{d\mathcal{A}}{d\nu}\right)\right\} \\
		&= \sqrt{2\lambda_h}v_0\nu\frac{dQ}{d\nu} +\underbrace{ \frac{4 \pi v_0}{\sqrt{2\lambda_h}} \int d \rho \rho^2 \left\{-\alpha^2 \mathcal{A}\frac{d}{d\nu}\left[(\nu-\alpha^2 \mathcal{A})\Phi^2\right]+\alpha^2\left(\partial_\rho \mathcal{A}\right)\left(\partial_\rho \frac{d\mathcal{A}}{d\nu}\right)\right\}}_{=0}\,\,.
	\end{aligned}
\end{equation}
In the third line we have integrated $\partial_\rho \mathcal{A}$ by part and used Eq.~\eqref{Ap}, then the integral vanishes. Finally,
\begin{equation}
	\frac{dE}{d\omega} = \frac{ dE}{d\nu} \frac{d\nu}{d\omega} = \left(\sqrt{2\lambda_h}v_0\omega\frac{dQ}{d\omega}\right)\left(\frac{1}{\sqrt{2\lambda_h}v_0}\right) = \omega\frac{dQ}{d\omega}\,\,,
\end{equation}
which leads to 
\begin{equation}
	\frac{dE}{dQ} = \omega
\end{equation}
for $\frac{dQ}{d\omega}\neq 0 $. The existence of $\frac{dQ}{d\omega}= 0$ indicates the locally minimal or locally maximal charge.

\begin{figure}[h]
\centering
	\begin{minipage}{1.0\linewidth}
		\vspace{3pt}
		\centerline{\includegraphics[width=1.1\textwidth]{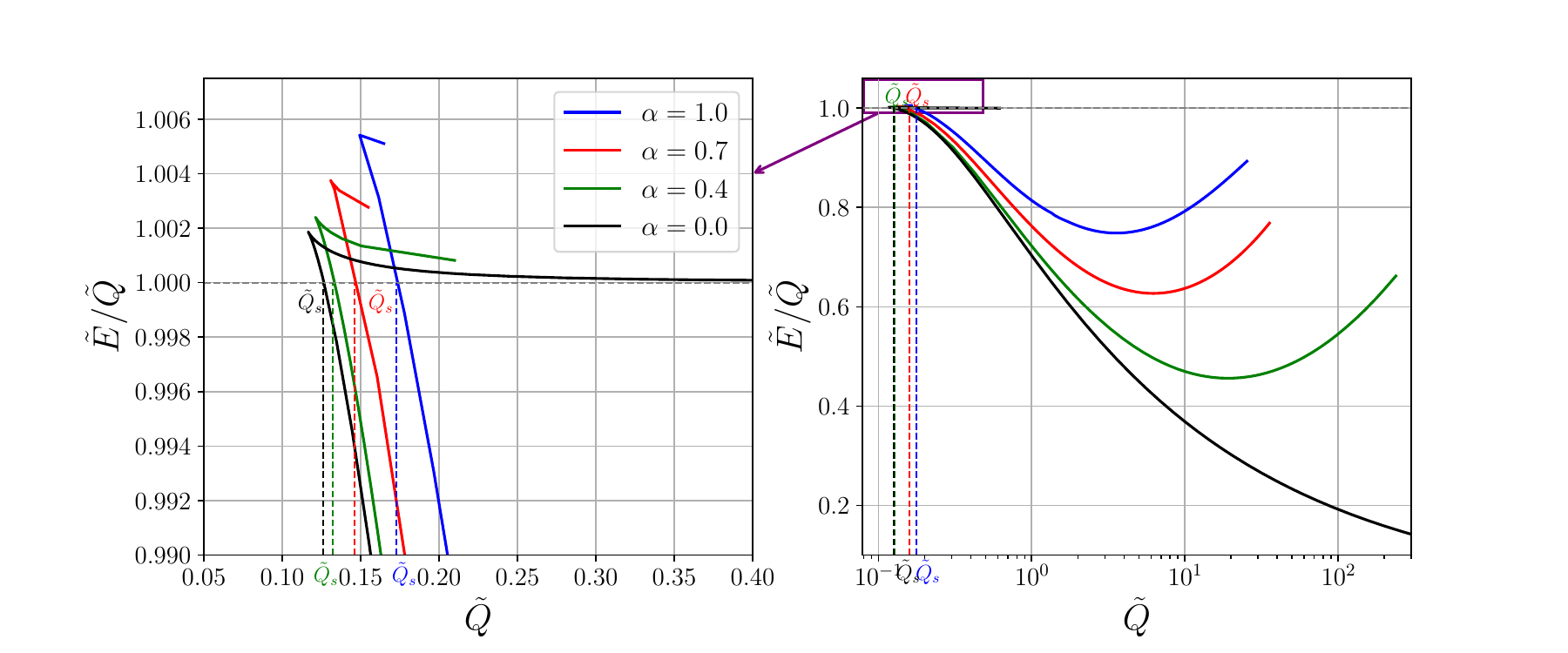}}
	\end{minipage}
	\caption{The energy over charge $\tilde E/\tilde Q$ for different values of $\alpha$. The $\tilde Q_s$ represents the value of charge which satisfies $\tilde E/\tilde Q=1$.}
	\label{EoverQ}
\end{figure}

\subsection{Stability of gauged Q-balls}
The stability of Q-balls is an important criterion to judge whether they can serve as the DM candidate. Unlike the global Q-balls, the stability of gauged Q-balls is still being discussed. In this subsection, we will systematically analyze four stability criteria of gauged Q-balls and show the viable parameter space of stable gauged Q-balls.
\subsubsection{Quantum mechanical stability}

The quantum mechanical stability is satisfied if 
\begin{equation}\label{quantum}
E<m_\phi Q	\quad \text{or} \quad  \tilde E/\tilde Q < 1\,\,.
\end{equation}
This means that the gauged Q-balls are stable against decay to free scalar particles. It should be noted that if the
Q-ball has decay channels into other fundamental scalar particles which have the mass $m_i$ that are smaller than $m_\phi$, we need to replace $m_\phi$ by $m_i$ in Eq.~\eqref{quantum}. The decay of global Q-balls or gauged Q-balls has been discussed in several works~\cite{Cohen:1986ct,Kawasaki:2012gk,Hong:2017uhi,Kasuya:2012mh}. When the effective energy of DM particles inside the gauged Q-balls $\pi/r_\star \propto ( \omega-\tilde{g} \tilde{A}_t )$ is larger than the masses of decay products, the decay process is kinetically allowed. If the Q-ball radius $r_\star$ is large enough or $( \omega-\tilde{g} \tilde{A}_t )$ is small, the gauged Q-balls do not decay into other daughter particles and thus are stable .

The ratio $\tilde E/\tilde Q$ is shown in the  figure \ref{EoverQ}. As the frequency $\nu \rightarrow k$ on the first branch, there is a region of parameter space where $\tilde E/\tilde Q>1$. It implies
the existence of a minimal Q-ball charge $\tilde Q_s$ defined as $\tilde E(Q_s)/\tilde Q_s=1$ of the quantum mechanically stable gauged Q-ball.
We can see from figure \ref{EoverQ} that for the global Q-balls the $\tilde E/\tilde Q$ decreases with growing $\tilde Q$ when $\tilde E/\tilde Q<1$. The branch of $\tilde E/\tilde Q>1$ corresponds to $\nu \rightarrow k$. So the global Q-balls are quantum mechanically stable as $\nu \ll k$ and $\tilde Q > \tilde Q_s$. One would wonder that the gauged Q-ball will destroy the quantum mechanical stability at large charge because the dark electrostatic energy is proportional to $\tilde Q^2$ and thus the $\tilde E/\tilde Q$ is proportional to $\tilde Q$. However, we found the gauged Q-balls are always quantum mechanically stable on the second branch where the gauge field dominates because of the charge of the gauged Q-balls must be finite.

\subsubsection{Stress stability}
Now we investigate the effects of the electrostatic repulsion on the stability of the gauged Q-balls. In Ref.~\cite{Loiko:2022noq}, the authors pointed out that, just like the hadrons, a necessary condition for stability of the configuration of gauged Q-balls is the balance of the internal forces, called von Laue condition~\cite{Laue:1911lrk,Bialynicki-Birula:1993shm}
\begin{equation}
	\int_{0}^{\infty} dr r^2 p(r) = 0\,\, . 
\end{equation}
Here  $p(r)$ is the radial distribution of the pressure inside the Q-ball, which can be extracted from the energy-momentum tensor by using the following parametrization~\cite{Polyakov:2002yz,Mai:2012yc}:
\begin{equation}
	T_{ij} = \left(\hat r_i \hat r_j - \frac{1}{3}\delta_{ij}\right)s(r)+\delta_{ij}p(r)\,\,.
\end{equation}
$s(r)$ is the traceless part which yields the anisotropy of pressure (shear forces). This kind of stability has also been studied for global Q-balls~\cite{Mai:2012yc,Polyakov:2018zvc}.

One stronger local criterion is that the normal force per unit area
acting on an infinitesimal area element at a distance $r$, must be directed outward~\cite{Perevalova:2016dln,Polyakov:2018zvc},
\begin{equation}\label{inequal}
	F(r) = \frac{2}{3}s(r)+p(r)>0\,\,.
\end{equation}
This is a necessary but not sufficient condition for stability.
By using the rescaled parameters, we have the expressions of $p(\rho)$ and $s(\rho)$ in the FLSM model:
\begin{equation}
	\begin{aligned}
		p(\rho) &= 2\lambda_hv_0^4\left[-\frac{1}{6}(\partial_\rho \Phi)^2  -\frac{1}{6}(\partial_\rho \mathcal{H})^2 + \frac{\alpha^2}{6}(\partial_\rho \mathcal{A})^2 + \frac{1}{2}(\nu - \alpha^2 \mathcal{A})^2\Phi^2 - \frac{k^2}{2}\Phi^2\mathcal{H}^2 - \frac{1}{8}(\mathcal{H}^2-1)^2 \right]\,\,, \\
		s(\rho) &=   2\lambda_hv_0^4\left[(\partial_\rho \Phi)^2  +(\partial_\rho \mathcal{H})^2 - \alpha^2(\partial_\rho \mathcal{A})^2 \right]\,\,,
	\end{aligned}
\end{equation}
from which we get 
\begin{equation}\label{Fr}
	F(\rho) = 2\lambda_hv_0^4\left[\frac{1}{2}(\partial_\rho \Phi)^2  +\frac{1}{2}(\partial_\rho \mathcal{H})^2 - \frac{\alpha^2}{2}(\partial_\rho \mathcal{A})^2 + \frac{1}{2}(\nu - \alpha^2 \mathcal{A})^2\Phi^2 - \frac{k^2}{2}\Phi^2\mathcal{H}^2 - \frac{1}{8}(\mathcal{H}^2-1)^2 \right]\,\,.
\end{equation}

We briefly discuss the stress stability of global Q-balls by taking 
$\alpha \rightarrow 0$ in Eq.~(3.10).
	For global Q-balls, we have 
	\begin{equation}
	F(\rho)=2 \lambda_h v_0^4\left[\frac{1}{2}\left(\partial_\rho \Phi\right)^2+\frac{1}{2}\left(\partial_\rho \mathcal{H}\right)^2+\frac{1}{2}\nu^2 \Phi^2-\frac{k^2}{2} \Phi^2 \mathcal{H}^2-\frac{1}{8}\left(\mathcal{H}^2-1\right)^2\right]\,\,,
	\end{equation}
then we can get 
	\begin{equation}\label{dF}
	\frac{1}{2\lambda_h v_0^4}\frac{d F(\rho)}{d\rho} = \partial_\rho\Phi \left[\partial_\rho^2 \Phi + \nu^2 \Phi - k^2 \Phi\mathcal{H}^2  \right] +\partial_\rho\mathcal{H}\left[\partial_\rho^2\mathcal{H} - k^2 \Phi^2 \mathcal{H} - \frac{1}{2}\mathcal{H}(\mathcal{H}^2-1) \right] \,\,.
	\end{equation}
The EoM for $\Phi$ and $\mathcal{H}$ in the FLSM model for $\alpha=0$ read
	\begin{equation}\label{EoM}
	\begin{aligned}
		&\frac{1}{\rho^2} \partial_\rho\left(\rho^2 \partial_\rho \Phi\right)+\left[\nu^2-k^2 \mathcal{H}^2\right] \Phi=0\,\,,\\
		&\frac{1}{\rho^2} \partial_\rho\left(\rho^2 \partial_\rho \mathcal{H}\right)-k^2 \mathcal{H} \Phi^2-\frac{1}{2} \mathcal{H}\left(\mathcal{H}^2-1\right)=0\,\,.
	\end{aligned} 
	\end{equation}
Substituting Eq.~\eqref{EoM} into Eq.~\eqref{dF}, 
    we obtain
    \begin{equation}
	\frac{1}{2\lambda_h v_0^4}\frac{d F(\rho)}{d\rho} = -\frac{2}{\rho}\left[(\partial_\rho \Phi)^2 + (\partial_\rho \mathcal{H})^2\right]\,\,.
	\end{equation}
This implies that $F(\rho)$ is a decreasing function of $\rho$. Because $F(\rho)$ must satisfy $F(\rho) \rightarrow 0 $ at large $\rho$, we find that $F(\rho)$ is always positive. 
Then the global Q-balls will be always stable 
under stress stability criterion. 
Based on this, it can also be expected that the gauged Q-balls are also stable under stress stability on the first branch where the backreaction 
from gauge field is small enough.

For gauged Q-balls, we can first consider one extreme case. 
At the end point where $( \nu - \alpha^2 \mathcal{A} ) \rightarrow 0$ 
on the second branch,  the $\Phi$ is finite
and $\partial_\rho \Phi = \partial_\rho \mathcal{H} =\partial_\rho \mathcal{A} =0$ at $\rho =0$. 
Then, from Eq.~\eqref{Fr}, we can see that
   \begin{equation}\label{F0}
   \frac{F(\rho)}{ 2\lambda_hv_0^4} \rightarrow -\frac{1}{8} \quad \text{at $\rho=0$}
   	\end{equation}
as $\mathcal{H} \approx 0$ inside gauged Q-balls in the large ball (thin-wall) limit. This implies the gauged Q-balls with maximal values of gauge filed $\mathcal{A} = \nu/\alpha^2$ inevitably break the stress stability.

\begin{figure}[h]
	\begin{minipage}{0.5\linewidth}
		\vspace{3pt}
		\centerline{\includegraphics[width=\textwidth]{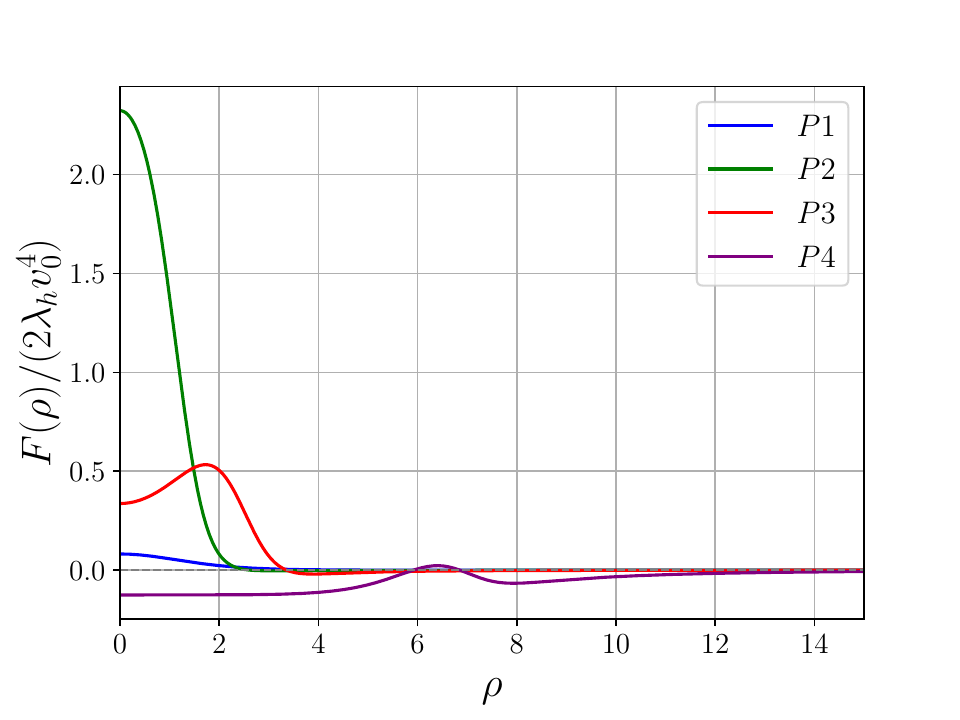}}
	\end{minipage}
	\begin{minipage}{0.5\linewidth}
		\vspace{3pt}
		\centerline{\includegraphics[width=\textwidth]{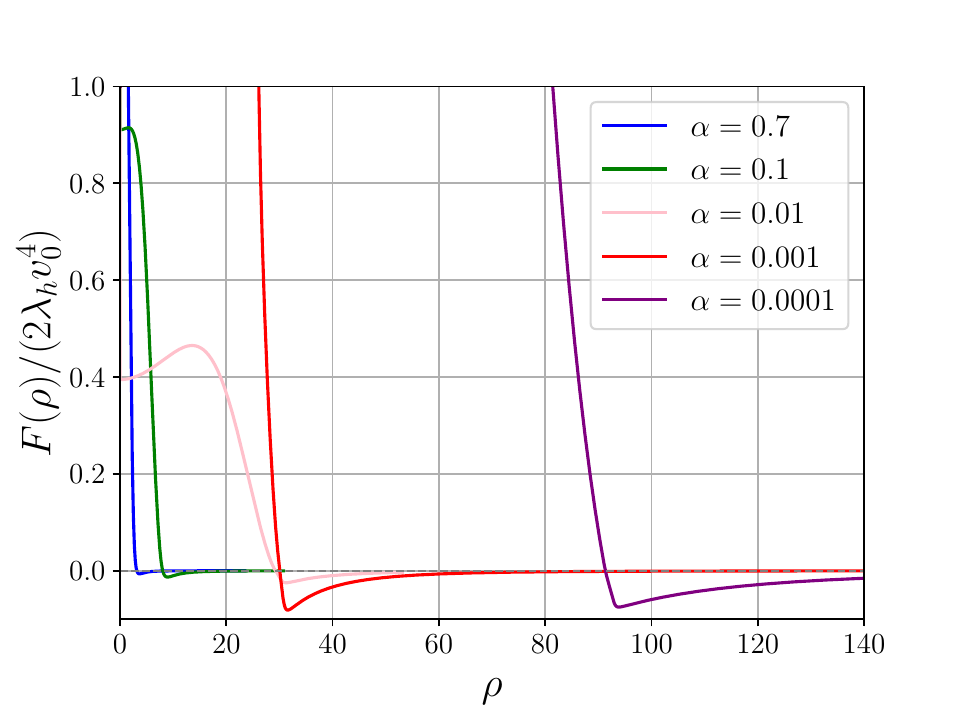}}
	\end{minipage}
	\caption{Left: the profile of $F(\rho)$ defined by Eq.~\eqref{Fr} for the four marked points in figure \ref{field0} where $\alpha=1.0$. Right:  $F(\rho)$ at the transition point between the first and second branches for different values of gauge couplings $\alpha$.}
	\label{stress}
\end{figure}

We also expect the gauged Q-balls are unstable under 
stress stability criterion on the second branch where the gauge field 
dominates. We prove this by using the numerical calculations.
In the left panel of figure \ref{stress}, we show the profile of $F(\rho)$ for four 
marked points in figure \ref{field0} where $\alpha=1.0$. We can see that the $F(\rho)$ has no nodes for $P1$ and $P2$ on the first branch. On the other hand, negative values exist for $F(\rho)$ on the second branch.
The gauged Q-balls on the second branch where the gauge field dominates are unstable. $F(\rho)$ behaves similarly to Eq.~\eqref{F0} at $\rho=0$ when $(\nu-\alpha^2 \mathcal{A} ) \rightarrow 0$, which can also be seen from the 
$P4$ of figure 5. It should be noted that the inequality \eqref{inequal} takes an approximation that Q-balls behave as a continuous media, which needs more discussions. And in the right panel we plot the $F(\rho)$ at the  transition point between the first and second branch for different values of gauge coupling $\alpha$. The values of $\alpha$ are set to be $0.0001-0.7$. It can be seen that they are all unstable under stress stability criterion. This implies that the solutions on the second branch are also unstable where the gauged field becomes larger~\cite{Loiko:2022noq}. The conclusion does not change qualitatively for different values of $\alpha$.

The stress-stability is the most stringent stability criterion in this work. We are not sure if the gauged Q-ball on the second branch would decay into free particles or smaller Q-balls. And it is meaningful to explore that how we can get rid of the strong constraints from stress-stability, in other words, how to get large gauged Q-balls with large gauge coupling. Maybe we can consider an example of two-scalar case where the two scalars $\phi$ and $\psi$ possess opposite charge. If the electrostatic fields produced by the two scalars cancels with each other, which implies
\begin{equation}
	\tilde{g}Q_\phi + \tilde{g}'Q_\psi=0\,\,,   
\end{equation}
with $\tilde{g}$ and $\tilde{g}'$ being the gauge couplings of $\phi$ and $\psi$ respectively, this guarantees the electric neutrality of the interior of the gauged Q-balls~\cite{Anagnostopoulos:2001dh,Ishihara:2018rxg}.
The gauged Q-balls will avoid the electric repulsion which leads to the stress-instability even when for large gauge coupling. We expect these Q-balls can also be the DM candidate because they will behave as the ordinary global Q-balls.

\subsubsection{Stability against fission}
For nongauged global Q-balls the corresponding stability criterion against fission takes the form
\begin{equation}
	d^2E/dQ^2<0 \,\,.
\end{equation}
This clearly leads to $E(Q_1)+E(Q_2)>E(Q_1+Q_2)$ when $E(0)=0$. However, it may not hold everywhere for the gauged Q-balls due to the presence of the gauge potential. In Ref.~\cite{Gulamov:2013cra} the authors gave a detailed discussion on the stability against fission of $U(1)$ gauged Q-balls. They pointed out that we could not make any conclusion about the stability
against fission for gauged Q-balls on the second branch. Nevertheless, it has been shown that the gauged Q-balls on the first branch are generally stable because the backreaction of the gauge field is generally small.

\subsubsection{Classical stability}

The problem of classical stability of $U(1)$ gauged Q-balls is discussed in detail in Ref.~\cite{Panin:2016ooo}. The classical stability criterion was firstly derived in \cite{Lee:1991ax} for one-field Q-balls and was discussed for the model with two scalar fields in \cite{Friedberg:1976me}.
The proof of Refs.~\cite{Friedberg:1976me,Lee:1991ax} was based on examining the properties of the energy functional of the
system. Instead, the examination of  Ref.~\cite{Panin:2016ooo} is based on the Vakhitov-Kolokolov method~\cite{1973Stationary,1978Dynamics} which utilized linearized EoM for the perturbations above the background solution. 

We only consider the spherical perturbations on the gauged Q-ball.
We adopt the following ansatz:
\begin{equation}
	\begin{aligned}
		\phi(t,r) &= e^{-i\omega t}f(r)+e^{-i\omega t}e^{\gamma t} (u(r)+il(r))\,\,, \\
		\tilde A_t(t,r) & = \tilde A_t(r)+e^{\gamma t}a_0(r)\,\,, \\
		h(t,r) &= h(r) + e^{\gamma t} \sigma(r)\,\,,
	\end{aligned}
\end{equation}
where $f(r)$, $\tilde A_t(r)$ and $h(r)$ are the background solutions. $u(r)$, $l(r)$, $a_0(r)$ and $\sigma(r)$ are the perturbations on the background.
Then we obtain the linearized EoM as below,
\begin{equation}\label{perturbations}
	\begin{aligned}
		&\Delta u +(\omega -\tilde{g}\tilde A_t)^2u - \gamma^2 u-2(\omega - \tilde{g}\tilde A_t)\gamma l - 2\tilde g(\omega-\tilde{g} \tilde A_t)fa_0 - Uu -2Su -Y\sigma = 0\,\,, \\
		&\Delta l+(\omega -\tilde{g} \tilde A_t)^2l - \gamma^2 l+2(\omega - \tilde{g} \tilde A_t)\gamma u -\tilde{g}\gamma fa_0 -Ul =0\,\,, \\
		&\Delta a_0 -2\tilde g^2f^2a_0 +4\tilde g(\omega-\tilde{g} \tilde A_t)fu-2\tilde g\gamma fl=0\,\,, \\
		& \Delta \sigma - \gamma^2 \sigma -W\sigma - 2Yu=0\,\,.
	\end{aligned}
\end{equation}
where $\Delta = \sum_{i=1}^3 \partial_i \partial_i$ is the 3-dim Laplacian operator and
\begin{equation}
	\begin{aligned}
		&U(r) = \left.\frac{\partial V}{\partial (\phi^{\dagger }\phi)}\right|_{\substack{\phi^\dagger \phi=f^2(r) \\ h=h(r)}}, \quad S(r) = \left.\frac{\partial^2 V}{\partial (\phi^{\dagger }\phi)^2}\right|_{\substack{\phi^\dagger  \phi=f^2(r) \\ h=h(r)}} f^2(r)\,\,, \\
		&W(r)=\left.\frac{1}{2} \frac{\partial^2 V}{\partial h^2}\right|_{\substack{\phi^\dagger  \phi=f^2(r) \\ h=h(r)}}, \quad Y(r)=\left.\frac{\partial^2 V}{\partial\left(\phi^\dagger  \phi\right) \partial h}\right|_{\substack{\phi^\dagger  \phi=f^2(r) \\ h=h(r)}} f(r)\,\,.
	\end{aligned}
\end{equation}
The boundary conditions are
\begin{equation}
	u'(0)=l'(0)=a_0'(0)=\sigma'(0)=0, \quad u(\infty)=l(\infty)=a_0(\infty)=\sigma(\infty)=0\,\,.
\end{equation}

We want to obtain the parameter $\gamma$ which depicts the growth of perturbations $u,l,a_0,\sigma$. The classically unstable mode corresponds to $\gamma>0$. This can be done by using the shooting method in Ref.~\cite{Levkov:2017paj}. We introduce four basis solutions $\Psi^{(i=1,2,3,4)} (r)= (u^{(i)},l^{(i)},a_0^{(i)},\sigma^{(i)})$ which satisfy the Neumann boundary conditions $d\Psi^{(i)}/dr|_{r=0}=0$ and Dirichlet boundary conditions $\Psi^{(1)} (0)=(1,0,0,0)$, $\Psi^{(2)} (0)=(0,1,0,0)$, $\Psi^{(3)} (0)=(0,0,1,0)$ and $\Psi^{(4)} (0)=(0,0,0,1)$. Then we integrate the Eqs.~\eqref{perturbations} numerically to find the values of $\Psi^{(i)}$ at large $r=r_\infty$. Now, recall that we are searching for a specific solution
$\Psi(r) = c_1 \Psi^{(1)}(r) +c_2 \Psi^{(2)}(r)+c_3 \Psi^{(3)}(r)+c_4 \Psi^{(4)}(r)$ which satisfies $\Psi(r_\infty) = 0$. This gives the system of linear equations,
\begin{equation}\label{matrixeq}
	\hat{D}_c (c_1,c_2,c_3,c_4)^{T}=0, \quad \hat{D}_c = (\Psi^{(1)T},\Psi^{(2)T},\Psi^{(3)T},\Psi^{(4)T})\,\,.
\end{equation}

\begin{figure}[h]
\centering
	\begin{minipage}{0.65\linewidth}
		\vspace{3pt}
		\centerline{\includegraphics[width=\textwidth]{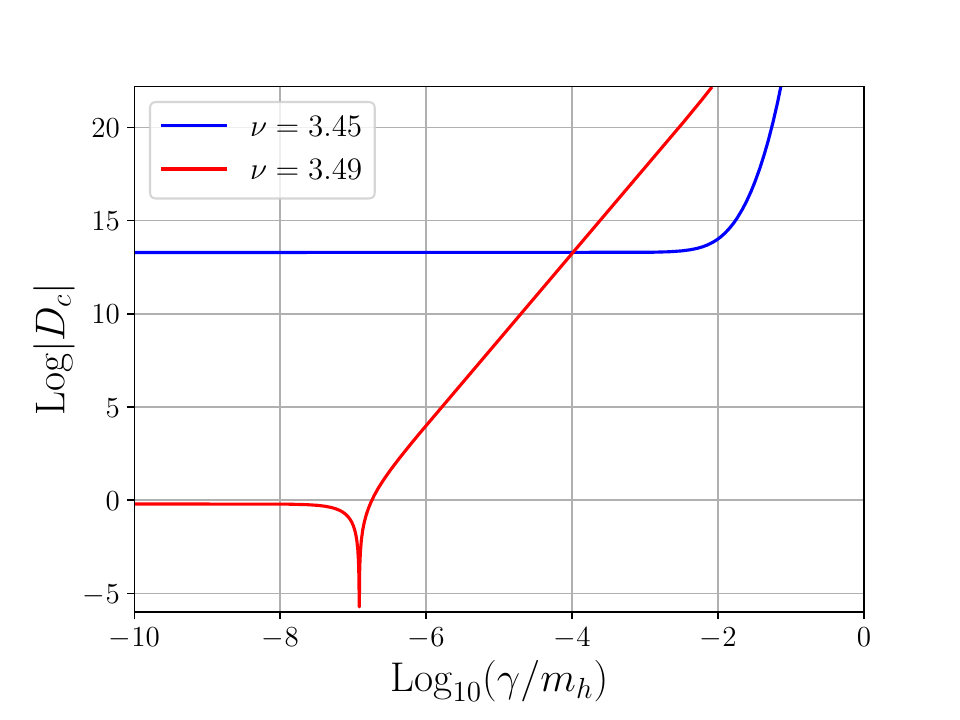}}
	\end{minipage}
	\caption{Determinant of matrix for $\nu=3.45$ and $\nu=3.49$ on the first branch respectively. We choose $\alpha^2=10^{-5}$, $k=3.5$. }
	\label{classical}
\end{figure}

Eq.~\eqref{matrixeq} has nontrivial solutions only if $D_c \equiv \det \hat{D}_c = 0$. In figure \ref{classical}
we plot $\mathrm{Log} |D_c|$ as a function of $\mathrm{Log}_{10}\gamma$ where we choose $\alpha^2=10^{-5}$ and $k=3.5$. We found that there is no classically unstable mode for $\nu = 3.45$ as there is no solution for $\gamma>0$. This also holds for other solutions of 
$U(1)$ gauged Q-balls except for $\nu \rightarrow k$ on the first branch. We found there is one unstable mode for $\nu=3.49$. Actually, in the limit $\nu \rightarrow k$, the contribution of gauged field can be neglected and one expects that the case is similar to the non-gauged Q-ball where $\frac{dQ}{d\omega}>0$ indicates there exists classically unstable mode. This is also discussed in  Refs.~\cite{Friedberg:1976me,Panin:2016ooo}. However, we usually do not have to worry about this because the region already has been excluded by the quantum instability. In Ref.~\cite{Kinach:2022jdx}, the authors have shown that the gauged Q-balls on the second branch with small gauge coupling are classically unstable with respect to  axisymmetric perturbations. This enhances our confidence that the gauged Q-ball on the second branch is unstable. The region is almost covered by the stress stability criterion.

\begin{figure}[h]
\centering
	\begin{minipage}{0.65\linewidth}
		\vspace{3pt}
		\centerline{\includegraphics[width=\textwidth]{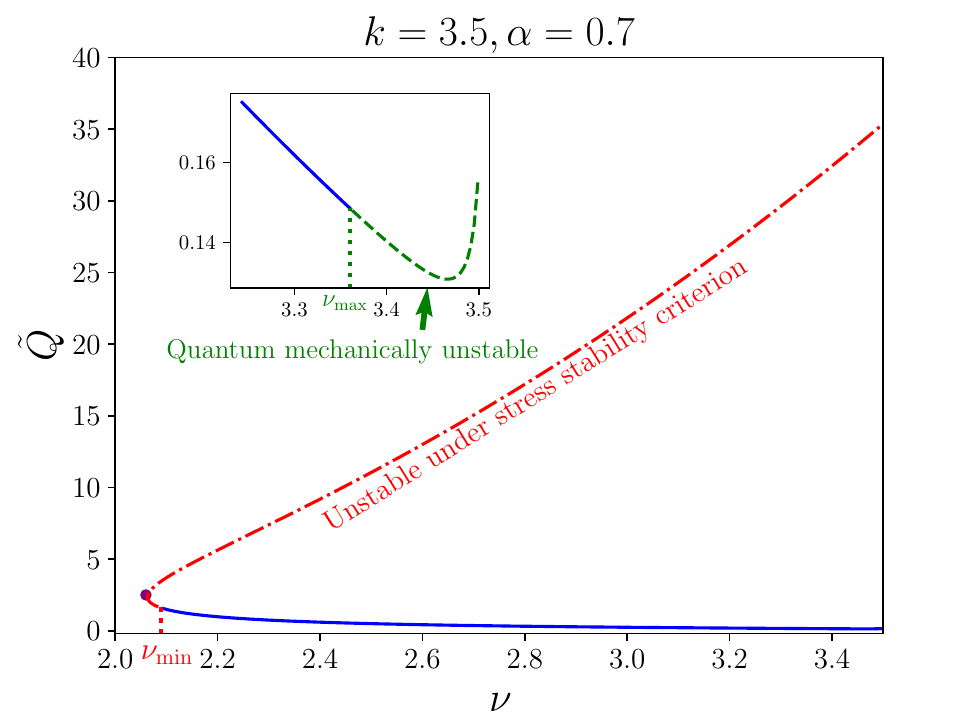}}
	\end{minipage}
	\caption{Viable space of $\nu$ of gauged Q-balls for $\alpha=0.7$. The red line represents the region where the gauged Q-ball is dominated by gauge field such that it is unstable under stress stability criterion. The green line represents the region where the gauged Q-ball is unstable under quantum stability criterion. The purple point labels the transition point. Gauged Q-balls in the blue region between $\nu_{\mathrm{min}}$ and $\nu_{\mathrm{max}}$ are stable.}
	\label{Qparam}
\end{figure}

In summary, the parameter space of gauged Q-balls is shown in figure \ref{Qparam}. The red line represents the region where the gauged Q-ball is dominated by the gauge field such that it is unstable under stress stability criterion. The green line represents the region where the gauged Q-ball is unstable under quantum stability criterion. We only plot the quantum mechanical stability and the stress stability criterion because they cover the space where the gauged Q-balls are classically unstable and are unstable against fission respectively. The gauged Q-balls are stable only in the region of $\nu \in [\nu_{\mathrm{min}},\nu_{\mathrm{max}}]$. The $\nu_{\mathrm{min}}$ corresponds to the maximal charge of gauged Q-balls and is close to the transition point between first and second branches which is marked by the purple point. In order to form gauged Q-balls of given charge at given gauge coupling, the charge has to be smaller than the maximal charge. It should also be noted that, if the dark gauge boson of gauged Q-ball kinetically mixes with SM
photons or $Z$ bosons, this would produce distinct experimental signatures which can constrain Q-ball
charge and couplings.

\section{Thin-wall approximation}\label{thinwall}
As the radii of gauged Q-balls become large, the width of  profile can be neglected and the gauged Q-balls can be depicted by the thin-wall approximation. The Higgs profile can be approximately viewed as a step function, the vacuum value is equal to zero and $v_0$ inside and outside the Q-balls, respectively. The derivative of the Higgs field only contributes to the surface term of gauged Q-balls which is negligible when the radius of the ball is large.
Then the problems are reduced to those or ones
with two fields, $\phi$ and $\tilde A_\mu$. We will discuss the simplified 
piecewise model and show that it behaves similarly to the FLSM model. By using the mapping method introduced by Ref.~\cite{Heeck:2021zvk}, we give some 
semi-analytic results and some analytic evaluations of the maximal charge of 
gauged Q-balls.
\subsection{Piecewise model}
If the Higgs field is approximately $h(\rho) = v_0\Theta(\rho-\rho_\star)$, then we can approximately view the complex scalar moving in the piecewise parabolic potential~\cite{Rosen:1968mfz,Gulamov:2013ema,Kim:2023zvf}. 
The Lagrangian density can be further approximated as
\begin{equation}\label{piecepotential}
\mathcal{L}_{\mathrm{piecewise}}=\left(D_\mu \phi\right)^{\dagger}\left(D^\mu \phi\right)-\frac{1}{4}  \tilde A_{\mu \nu} \tilde A^{\mu \nu}-m_\phi^2 \phi^\dagger\phi \Theta\left(1-\frac{\phi^\dagger\phi}{v^2}\right)-m_\phi^2v^2\Theta\left(\frac{\phi^\dagger\phi}{v^2}-1\right)\,\,, 
\end{equation}
where $\Theta(x)$ is the Heaviside step function. In our case,
\begin{equation}
	m_\phi = v_0 \sqrt{\frac{\lambda_{\phi h}}{2}}, \quad v = v_0\sqrt{\frac{\lambda_{ h}}{2\lambda_{\phi h}}}\,\,.
\end{equation} 
Note that $v$ is chosen so that $V(\phi)$ is continuous at $\phi^\dagger\phi = v^2$.
This can be understood from the EoM of the Higgs field in the FLSM model,
\begin{equation}
	h''(r) + \frac{2}{r}h'(r) + \left[\frac{m_h^2}{2} - \lambda_h h(r)^2 -  \lambda_{\phi h} f(r)^2\right] h(r) =0\,\,,
\end{equation}
where we use the definition $\phi(r,t) = f(r)e^{-i\omega t}$ and the prime denotes a derivative with respect to $r$. If the Higgs field 
is approximately a step function, we can neglect the derivatives, then we have
\begin{equation}
h^2 \approx	\begin{cases}
		\frac{m_h^2}{2\lambda_h} - \frac{\lambda_{\phi h}}{\lambda_h}f^2 \quad \text{for} \quad 2\lambda_{\phi h} f^2 < m_h^2 \,\,, \\
		0 \quad \text{for} \quad 2\lambda_{\phi h} f^2 > m_h^2 \,\,,
	\end{cases}
\end{equation}
and we could assume $f(r) \approx 0$ outside the bubble where $h^2 \approx \frac{m_h^2}{2\lambda_h} = v_0^2$. Therefore $V(\phi,h) \approx m_\phi^2 \phi^\dagger \phi$ and $V(\phi,h) \approx \frac{\lambda_h}{4}v_0^4$ outside and inside the Q-ball, respectively. The consistency between the piecewise model and the Friedberg-Lee-Sirlin model has been discussed in Ref.~\cite{Kim:2023zvf} and the classical stability of gauged Q-balls in piecewise model has been studied in Ref.~\cite{Panin:2016ooo,Gulamov:2013ema}.
The EoM of gauged Q-balls in the piecewise model after the rescaling 
of Eq.~\eqref{rescal} are
\begin{equation}\label{Ap2}
	\frac{1}{\rho^2} \partial_\rho\left(\rho^2 \partial_\rho \mathcal{A}\right)+ (\nu-\alpha^2 \mathcal{A}) \Phi^2=0\,\,,
\end{equation}
\begin{equation}\label{Bp2}
	\frac{1}{\rho^2} \partial_\rho\left(\rho^2 \partial_\rho \Phi\right)+\left[(\nu-\alpha^2 \mathcal{A})^2-k^2 \Theta\left(1-4k^2\Phi^2\right)\right] \Phi=0\,\,.
\end{equation}
These equations are easier to solve than the FLSM model by using undershooting/overshooting method. The gauged Q-ball energy and charge reads,
\begin{equation}\label{EQpiece}
\begin{aligned}
	\tilde{E} =\frac{1}{k} \int_0^{\infty} d \rho \rho^2 \mathcal{E}, \quad
	\tilde{Q} =\int_0^{\infty} d \rho \rho^2 (\nu-\alpha^2 \mathcal{A}) \Phi^2\,\,.
\end{aligned}
\end{equation}
Here, $\mathcal{E}=\frac{\alpha^2}{2 }\left(\partial_\rho \mathcal{A}\right)^2+\frac{1}{2}\left(\partial_\rho \Phi\right)^2+\frac{1}{2}\left[(\nu-\alpha^2 \mathcal{A})^2+k^2 \Theta(1-4k^2\Phi^2)\right] \Phi^2+\frac{1}{8}\Theta(4k^2\Phi^2-1)$ which has a different form from the 
FLSM model. 

We solve Eqs.~\eqref{Ap2} and \eqref{Bp2} numerically to get the profiles of 
the complex field and the gauge field. Then we can get the charge and energy of gauged Q-balls by substituting them into Eqs.~\eqref{EQpiece}. These numerical results are shown in figure \ref{EQR}. We can see that the piecewise model fits 
the FLSM model well when the gauged Q-ball radius is large. The distinctions 
appear at small $\rho_\star$ at which the Higgs field value inside is not approximately zero.

\subsection{Mapping gauged Q-balls}
We make a further assumption that the complex scalar field can also be viewed as a step function, $\Phi(\rho) = \Phi_0(1-\Theta(\rho-\rho_b))$, where we denote $\Phi_0=\Phi(0)$ and $\rho_b$ is defined by $\Phi(\rho_b)=\Phi_0/2$. Then the profile of gauged field is~\cite{Lee:1988ag}
\begin{equation}\label{Aana}
\mathcal{A}(\rho) = \frac{\nu}{\alpha^2}	
\begin{cases}
		1 - \frac{\sinh(\alpha \Phi_0 \rho)}{\cosh(\alpha \Phi_0\rho_b)\alpha \Phi_0 \rho}, & \rho<\rho_b \\
		\frac{\alpha \Phi_0\rho_b-\tanh(\alpha \Phi_0 \rho_b)}{\alpha \Phi_0 \rho }, & \rho>\rho_b\,\,.
	\end{cases}
\end{equation}
The Q-ball radius is defined by $\Phi(\rho_\star) = \frac{1}{2k}$.

\begin{figure*} [t!]
	\centering
	\subfigure{
		\includegraphics[scale=0.45]{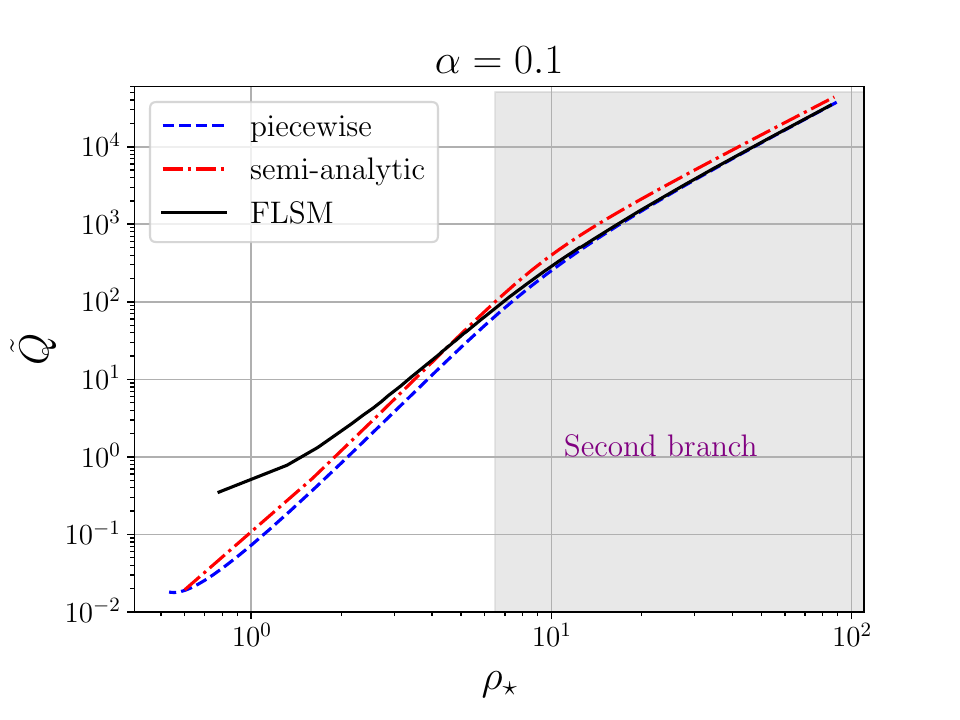}}
	\subfigure{
		\includegraphics[scale=0.45]{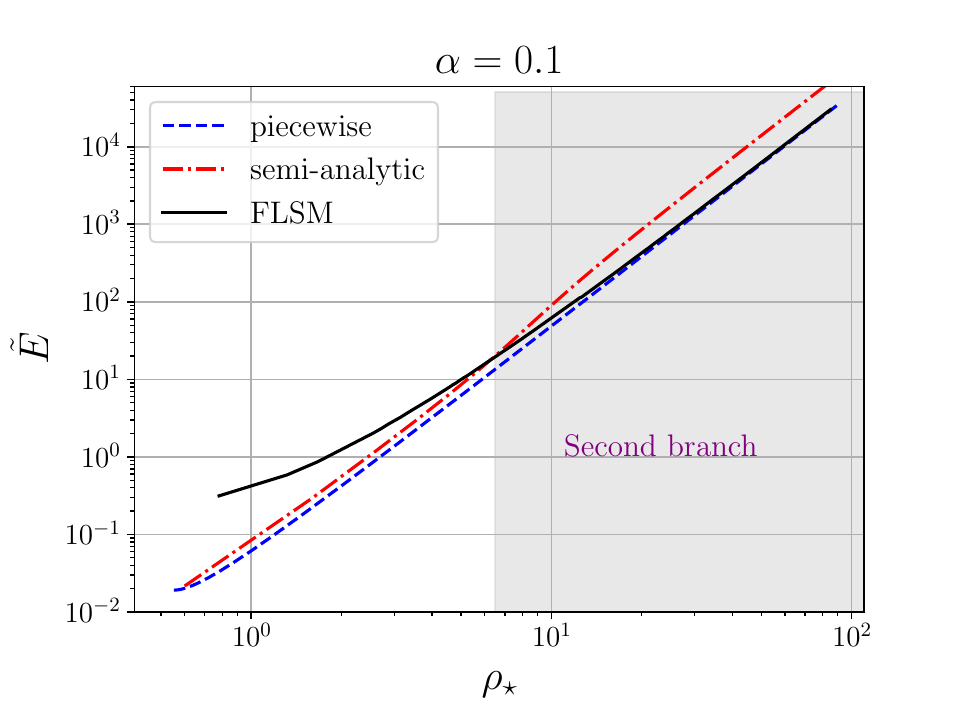}}
	\\
	\subfigure{
		\includegraphics[scale=0.45]{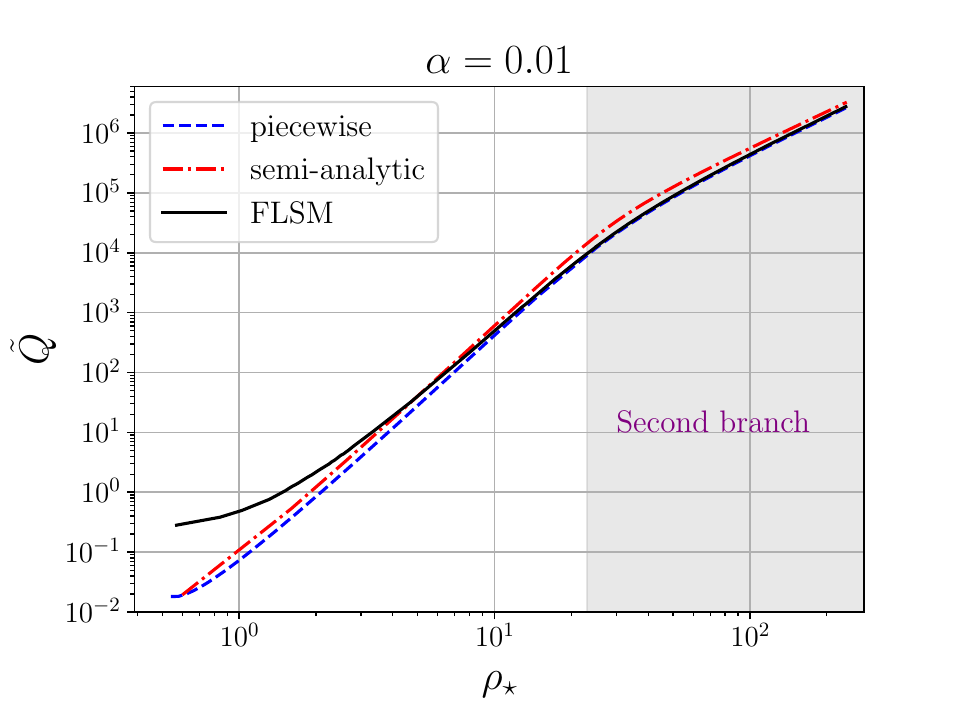}}
	\subfigure{
		\includegraphics[scale=0.45]{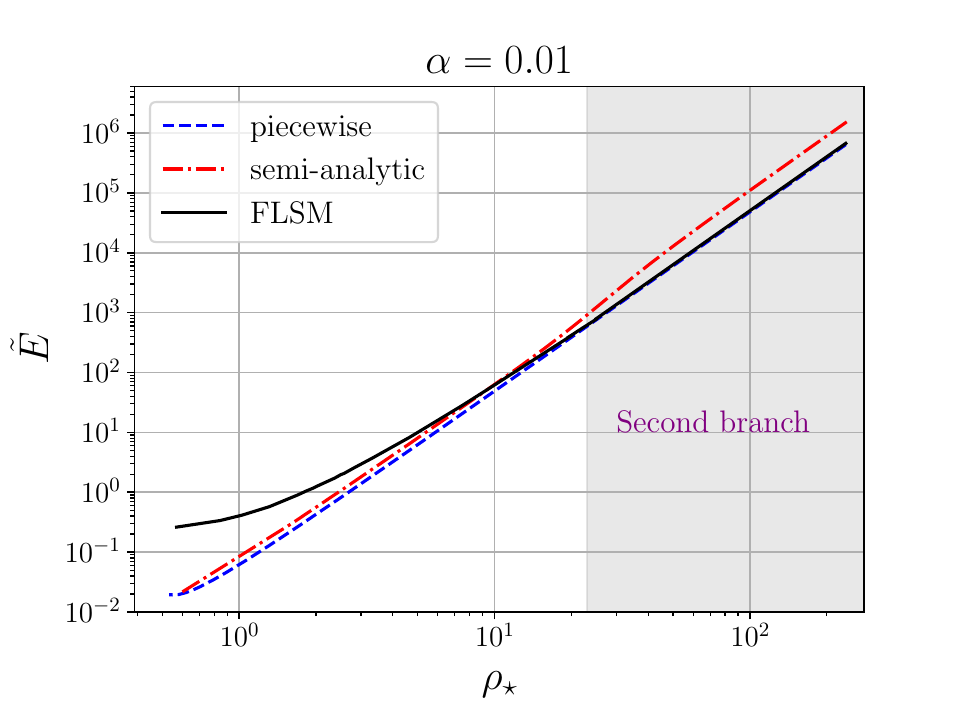}}
	\caption{Charge and Energy as functions of Gauged Q-ball radius. The blue lines are the numerical results in piecewise model; red lines are the semi-analytic results Eq.~\eqref{Qana} and Eq.~\eqref{Eana}, black lines come from the numerical results in FLSM model.}
	\label{EQR}
\end{figure*}

In Ref.~\cite{Heeck:2021zvk}, the authors propose a mapping between the gauged Q-ball and the global Q-ball. Specifically,
\begin{equation}\label{mapping}
	\nu = \nu_g \alpha \Phi_0 \rho_b \coth(\alpha \Phi_0 \rho_b)\,\,,
\end{equation}
where $\nu_g$ is the value of frequency for the global Q-ball with the same $\rho_b$. Then the profile of $\mathcal{A}$ is given by Eq.~\eqref{Aana}. This relation holds even for cases beyond the thin-wall approximation.  Interestingly, the global cases in the piecewise model have analytic solutions~\cite{Gulamov:2013ema}:
\begin{equation}\label{Bana}
	\Phi_g(\rho) = 
	\begin{cases}
		\frac{1}{2k}\frac{\rho_\star \sin(\nu_g \rho)}{\rho \sin(\nu_g \rho_\star)}, & \rho<\rho_\star \\
		\frac{1}{2k}\frac{\rho_\star e^{-\sqrt{k^2-\nu_g^2}\rho}}{\rho e^{-\sqrt{k^2-\nu_g^2}\rho_\star}}, & \rho>\rho_\star\,\,.
	\end{cases}
\end{equation}
This gives us $\Phi_0 = \frac{1}{2k}\frac{\nu_g\rho_\star}{\sin(\nu_g \rho_\star)}$ and $2\sin(\nu_g \rho_b) = \nu_g \rho_b$. Then we have $\nu_g \rho_b = C_1$ where $C_1 \approx 1.89549$\footnote{The factor $C_1$ is close to the result of Ref.~\cite{Heeck:2023idx} where $C_1 \approx 2.08$ by using the definition $\Phi''(\rho_b)=0$.}. The Q-ball radius is defined as
\begin{equation}\label{rhostar}
	\rho_\star(\nu_g) = \frac{1}{\nu_g} \left(\pi - \arctan \left(\frac{\nu_g}{\sqrt{k^2-\nu_g^2}}\right)\right)\,\,.
\end{equation}

If we use $\Phi(\rho) = \Phi_0(1-\Theta(\rho-\rho_b))$ and $\mathcal{A}(\rho)$ from Eq.~\eqref{Aana}, we have further semi-analytic results for charge:
\begin{equation}\label{Qana}
	\tilde Q = \frac{\nu_g \rho_b}{\alpha^2}\left(\alpha \Phi_0\rho_b\coth(\alpha \Phi_0 \rho_b)-1\right) = \frac{\rho_b}{\alpha^2}(\nu-\nu_g)\,\,.
\end{equation}
In the limit $\alpha \rightarrow 0$ and $\alpha \Phi_0\rho_b \rightarrow 0$, because $x\coth x \sim 1+\frac{x^2}{3} +\mathcal{O} (x^3)$ for  $x \rightarrow 0$. Then we have $\tilde Q \sim \nu_g \Phi_0^2 \rho_b^3$. And in this case, when $\nu_g \rightarrow 0$, from Eq.~\eqref{rhostar} we have $\nu_g \rho_\star \simeq \pi$ and $\sin(\nu_g \rho_\star) \simeq \nu_g/k$, which lead to
\begin{equation}\label{B0}
	\Phi_0 \simeq \frac{1}{2k} \frac{\pi}{\nu_g/k} = \frac{\pi}{2\nu_g}\,\,.
\end{equation}
From $2\sin(\nu_g \rho_b) = \nu_g \rho_b$, we have 
$\nu_g \sim \frac{C_1}{\rho_b}$, and
\begin{equation}
	\tilde Q \propto \rho_b^4\,\,,
\end{equation}
which is consistent with the global Q-ball case.
Using the same procedure that derives Eq.~\eqref{engQ}, we have analytic results for total energy:
\begin{equation}\label{Eana}
	\begin{aligned}
	\tilde E &= \frac{\nu \tilde Q}{k} + \frac{1}{3k} \int d \rho \rho^2\left[\left(\partial_\rho \Phi\right)^2-{\alpha^2}\left(\partial_\rho \mathcal{A}\right)^2\right]  \\
	& \simeq \frac{\nu \tilde Q}{k} + \frac{\Phi_0\rho_b^2}{12k} - \frac{1}{3k} \frac{\nu^2\left[\alpha \Phi_0 \rho_b(2+\mathrm{sech}^2(\alpha \Phi_0 \rho_b)) - 3\tanh (\alpha \Phi_0 \rho_b)\right] }{2\alpha^3\Phi_0}\,\,.
	\end{aligned}
\end{equation}
The second term comes from the integration over discontinuous $(\partial_\rho \Phi)^2$ by using the approximation of energy conservation~\cite{Heeck:2020bau}. 



In the first limit of $\alpha \Phi_0 \rho_b \rightarrow 0$ and large $\rho_b$, the last term of Eq.~\eqref{Eana} vanishes. We then have $\nu \simeq \nu_g$, $\Phi_0 \simeq \frac{\pi}{2\nu_g} \simeq \frac{\rho_\star}{2}$ and $\rho_\star \simeq \rho_b$, then
\begin{equation}\label{abrm1}
	\tilde E \simeq \frac{\pi \tilde Q}{k\rho_b} + \frac{\rho_b^3}{24k}\,\,,
\end{equation}
which is just the energy of global Q-ball.

In the opposite limit $\alpha \Phi_0 \rho_b \rightarrow \infty$, because $\mathcal{A} \rightarrow \tilde Q/\rho$, then from Eq.~\eqref{Aana},
\begin{equation}
	\nu = \frac{\alpha^2 \tilde Q}{\rho_b} \left(1-\frac{\tanh(\alpha \Phi_0 \rho_b)}{\alpha \Phi_0 \rho_b}\right)^{-1} \simeq \frac{\alpha^2 \tilde Q}{\rho_b}\,\,.
\end{equation}
Thus from Eq.~\eqref{Eana} the energy of gauged Q-ball is
\begin{equation}\label{abrl1}
	\begin{aligned}
\tilde E \simeq	\frac{\nu \tilde Q}{k} + \frac{\Phi_0\rho_b^2}{12k} - \frac{1}{3k}\frac{\nu^2 \rho_b}{\alpha^2} \simeq \frac{2\alpha^2 \tilde Q^2}{3k\rho_b} +  \frac{\Phi_0\rho_b^2}{12k}\,\,.
\end{aligned}
\end{equation} 
The first term is the Coulomb energy and the second term is the potential energy difference between inside and outside of the Q-ball.
The second term is proportional to $\rho_b^2 $ because in this case the Compton wavelength of the gauge field $\frac{1}{\tilde{g}v_0\Phi_0}$ inside the Q-ball is much smaller than Q-ball radius, $r_b = \rho_b/m_h$.
So the Q ball is superconducting. The potential
energy is therefore zero inside as well as outside of the Q
ball and is nonzero only in the shell around $\rho_b$~\cite{Lee:1988ag}.

For a given $\rho_\star$, we can solve the Eq.~\eqref{rhostar} to get the corresponding $\nu_g$. After using $\Phi_0 = \frac{1}{2k}\frac{\nu_g\rho_\star}{\sin(\nu_g \rho_\star)}$, $2\sin(\nu_g \rho_b) = \nu_g \rho_b$ and Eq.~\eqref{mapping} we can get $\Phi_0$, $\rho_b$ and $\nu$ of gauged Q-balls. Substituting them into
Eqs.~\eqref{Qana} and \eqref{Eana} gives the charge and energy of gauged Q-balls. These semi-analytic results are shown by the red lines in figure \ref{EQR}. We find the mapping works well for $\rho_\star \gg 1$ on the first branch where the profiles of scalar fields $\phi$ and $h$ can be safely viewed as step functions. However, the semi-analytic results of energy do not work well for the second branch because  $\Phi_0 \neq  \frac{1}{2k}\frac{\nu_g\rho_\star}{\sin(\nu_g \rho_\star)}$ in this case. The value of $\Phi_0$ is lower because the gauge potential dominates. This can be seen in figure \ref{field0}. At the transition point between the first and the second branches, the discrepancy between semi-analytic results and numerical results is about a factor of $\mathcal{O}(1)$.

\subsection{Maximal charge and energy of gauged Q-balls: analytic approximations}

We give analytic evaluations of maximal charge and maximal energy of gauged Q-balls for a given value of gauge coupling. Because the gauged Q-balls should be unstable under stress stability criterion on the second branch where the gauge potential dominates and the $\nu_{\mathrm{min}}$ is very close to the transition point, the maximal charge is approximately defined by $\left. \frac{d \nu}{d\rho_\star}\right|_{\rho_\star = \rho_{\mathrm{max}}}=0$. In the limit at large Q-ball with $\nu_g \rightarrow 0$, we have $\rho_b = C_1/\nu_g$ and $\Phi_0 \simeq  \frac{\rho_\star}{2}$. 
Then, from \eqref{mapping} we get
\begin{equation}
	\nu = \frac{C_1 \alpha \rho_\star}{2} \coth \left(\frac{C_1 \alpha \rho_\star}{2\nu_g}\right) = \frac{C_1 \alpha \rho_\star}{2} \coth \left(\frac{C_1 \alpha \rho_\star^2}{2\pi}\right)\,\,,
\end{equation}
where we used $\rho_\star \simeq \pi/\nu_g$ when $\nu_g \rightarrow 0$. 
Then from $\left. \frac{d \nu}{d\rho_\star}\right|_{\rho_\star = \rho_{\mathrm{max}}}=0$ when the charge is maximal, we have
\begin{equation}
	\sinh\left(\frac{C_1 \alpha \rho_{\mathrm{max}}^2}{2\pi}\right) 	\cosh\left(\frac{C_1 \alpha \rho_{\mathrm{max}}^2}{2\pi}\right) = \frac{C_1 \alpha \rho_{\mathrm{max}}^2}{\pi}\,\,,
\end{equation}
then we can obtain $\frac{C_1 \alpha \rho_{\mathrm{max}}^2}{2\pi} = C_2$ with $C_2 \approx 1.08866$ and $\alpha \Phi_0 \rho_b \simeq C_2$ which is somewhere in the middle of cases of Eq.~\eqref{abrm1} and Eq.~\eqref{abrl1}. We can also get the minimal frequency 
\begin{equation}
\nu_{\mathrm{min}} = \frac{C_1 \alpha \rho_{\mathrm{max}}}{2} \coth\left(\frac{C_1 \alpha \rho_{\mathrm{max}}^2}{2\pi}\right)= \sqrt{\frac{\pi C_1 C_2 \alpha}{2}} \coth(C_2) \approx 2.26 \sqrt{\alpha}\,\,,
\end{equation}
and the corresponding frequency for the global case $\nu_{g\mathrm{min}} = \frac{\pi}{\rho_{\mathrm{max}}} = \sqrt{\frac{\pi C_1 \alpha}{2C_2}}$. We can see that $\frac{\nu_{\mathrm{min}}}{\nu_{g\mathrm{min}}} = C_2\coth(C_2) \approx 1.367$ which is independent of the Q-ball size. This implies that as the solutions are unstable on the second branch, the gauged Q-ball lives on the first branch and is close to the global case. 

\begin{figure}[h]
	\begin{minipage}{0.48\linewidth}
		\vspace{3pt}
		\centerline{\includegraphics[width=\textwidth]{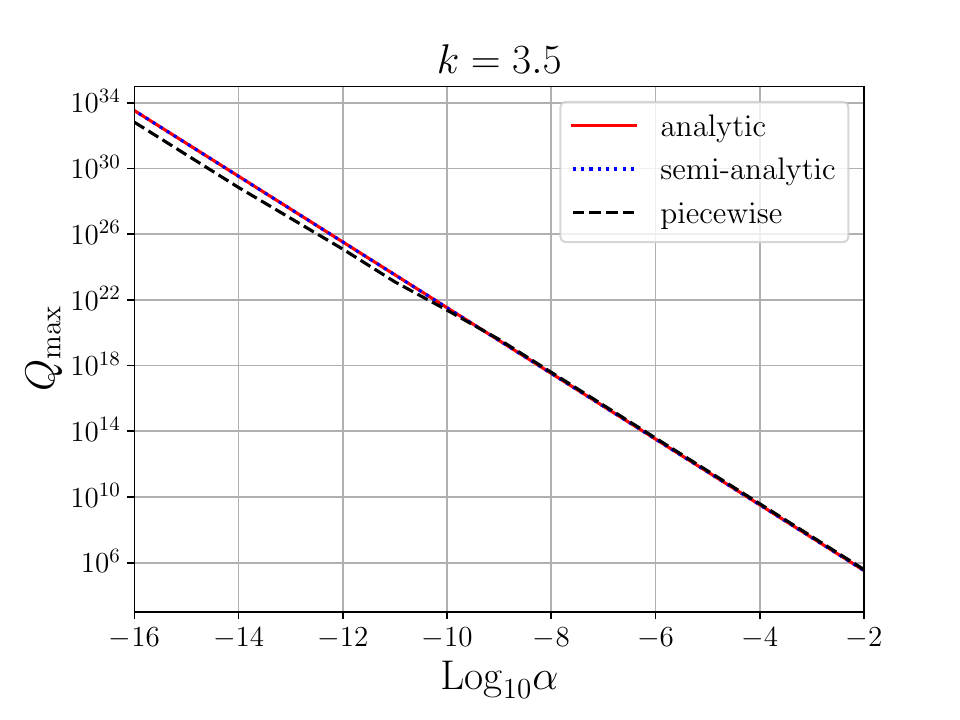}}
	\end{minipage}
	\begin{minipage}{0.48\linewidth}
		\vspace{3pt}
		\centerline{\includegraphics[width=\textwidth]{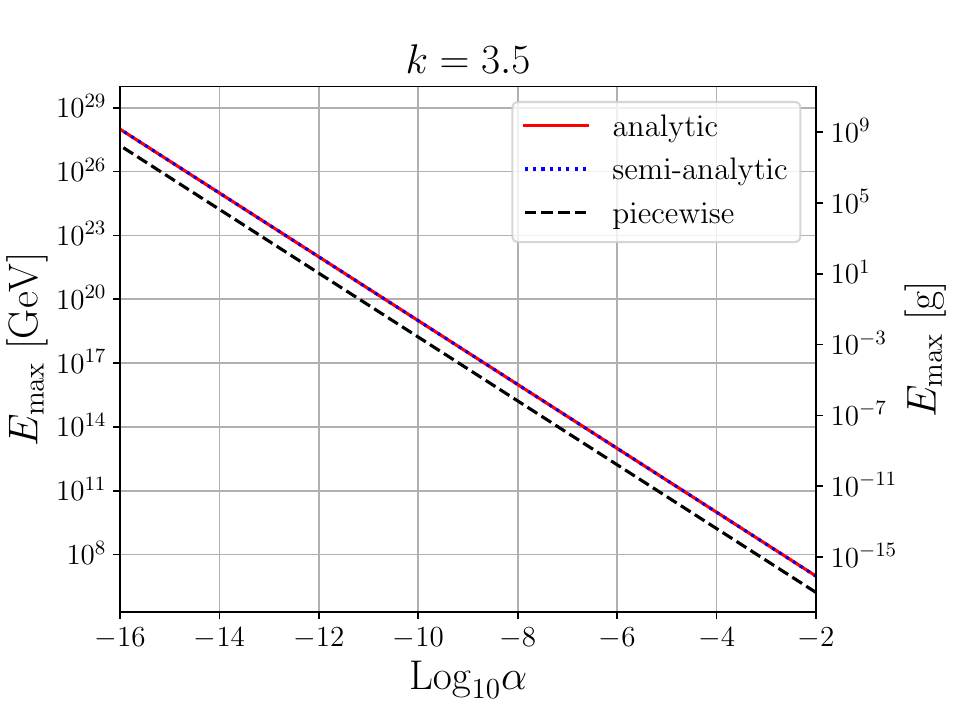}}
	\end{minipage}
	\caption{Maximal charge $Q_{\mathrm{max}} = \frac{2\pi}{\lambda_h}\tilde Q_{\mathrm{max}}$ (left panel) and energy $E_{\mathrm{max}} = \frac{2\pi m_\phi}{\lambda_h}\tilde E_{\mathrm{max}}$ (right panel) of gauged Q-ball for different $\alpha$ and $k$. The red lines are analytic evaluations Eq.~\eqref{Qanamax} and \eqref{Eanamax}; the blue dotted lines are the semi-analytic results and the black dashed lines represent the numerical results for piecewise model.
    }
	\label{Qmax}
\end{figure}

Finally, the maximal charge is given by
\begin{equation}\label{Qanamax}
\tilde Q_{\mathrm{max}} = \frac{C_1}{\alpha^2}(C_2\coth C_2-1) \approx 0.7 \alpha^{-2}\,\,,
\end{equation}
from which we can see the charge is unbounded from above in the global case where $\alpha=0$.
The maximal energy reads 
\begin{equation}\label{Emax}
	\begin{aligned}
	\tilde E_{\mathrm{max}} &= \frac{\nu_{\mathrm{min}}\tilde Q_{\mathrm{max}}}{k} + \frac{C_1^2 \rho_{\mathrm{max}}^3}{24\pi^2k}-\frac{1}{3k}\frac{\nu_{\mathrm{min}}^2[C_2(2+\mathrm{sech}^2C_2)-3\tanh C_2]}{\alpha^3 \rho_{\mathrm{max}}} \\
	&= \frac{\pi C_2 \coth C_2 \tilde Q_{\mathrm{max}} }{k\rho_{\mathrm{max}}}+ \frac{C_1^2 \rho_{\mathrm{max}}^3}{24\pi^2k}-\frac{1}{3k}\frac{\pi C_1C_2 \coth^2 C_2[C_2(2+\mathrm{sech}^2C_2)-3\tanh C_2]}{2\alpha^2 \rho_{\mathrm{max}}}\,\,.
	\end{aligned}
\end{equation}
By using $\nu_{\mathrm{min}}= \sqrt{\frac{\pi C_1 C_2 \alpha}{2}} \coth C_2$ and $\rho_{\mathrm{max}}= \sqrt{\frac{2\pi C_2}{C_1 \alpha}}$, we have
\begin{equation}\label{Eanamax}
	\begin{aligned}
	\tilde E_{\mathrm{max}}
	&=\frac{1}{k\alpha^{3/2}}\sqrt{\frac{\pi C_1 C_2}{72}}\left[-3C_1 \coth C_2+C_2\left(\frac{1}{\pi}+4C_1+3C_1 \mathrm{csch}^2 C_2\right)\right]\\
	&\approx 1.51 k^{-1} \alpha^{-3/2}\,\,.
	\end{aligned}
\end{equation}
This also gives us $\tilde E_{\mathrm{max}} \propto \rho_{\mathrm{max}}^3 \propto \tilde Q_{\mathrm{max}}^{3/4}$ which is expected in the global case. 

The analytic results of $Q_{\mathrm{max}} = \frac{2\pi}{\lambda_h}\tilde Q_{\mathrm{max}}$ and $E_{\mathrm{max}} = \frac{2\pi m_\phi}{\lambda_h}\tilde E_{\mathrm{max}}$ are shown in terms of the red lines in figure \ref{Qmax} and we can see that the maximal charge $Q_{\mathrm{max}}$ fits well with the semi-analytic and numerical results.  However, the analytic and semi-analytic $E_{\mathrm{max}}$ is about 5-6 times larger than the numerical results due to the uncertainties of values of $\Phi_0$.

\section{Gauged Q-ball DM from electroweak FOPT}\label{FOPTDM}
In the above discussions, we have shown that the gauged Q-ball could be stable and hence  be a  DM candidate. In this section, we begin to discuss the detailed production mechanism of
the gauged Q-ball DM from the electroweak FOPT in the early Universe. We consider the minimal Higgs extended model with a singlet scalar field which could trigger a strong FOPT~\cite{Espinosa:1993bs,Ponton:2019hux,Bandyopadhyay:2021ipw}. The discussions can also be applied to other FOPT models. The phase transition dynamics can be modified by introducing some new degrees of freedom beyond the standard model.

\subsection{Electroweak FOPT}

The electroweak FOPT dynamics is determined by the finite temperature effective potential $V_{\text {eff }}(h, T)$ where $h$ is the real component of the SM Higgs doublet as defined in Eq.~\eqref{zerolag},
\begin{equation}
V_{\text {eff }}(h, T) \equiv V_{\text {tree }}(h)+V_{\mathrm{CW}}\left(h\right)+V_{\mathrm{T}}\left(h, T\right)\,\,.
\end{equation}
The first term $V_{\text {tree }}(h)=\lambda_h\left(h^2-v_0^2\right)^2 / 4$ is the tree-level SM Higgs potential. $V_{\mathrm{CW}}(h)$ is the one-loop quantum correction to the effective potential, i.e., Coleman-Weinberg potential~\cite{Coleman:1973jx}. Using the on-shell renormalization scheme, we have
\begin{equation} V_{\mathrm{CW}}(h)=\sum_i(-1)^{F_i} \frac{g_i}{64 \pi^2}\left[m_i^4(h)\left(\log \frac{m_i^2(h)}{m_i^2(v_0)}-\frac{3}{2}\right)+2 m_i^2(h) m_i^2(v_0)\right],
\end{equation}
where $g_i$ is the degree of freedom for each particle, $F_i=1(0)$ for fermions(bosons), $m_i(h)$ are masses for $=t, W, Z, h, \phi$. The finite-temperature correction term is given by
\begin{equation}
 V_{\mathrm{T}}(h,T)=\sum_i(-1)^{F_i} \frac{g_i T^4}{2 \pi^2} \int_0^{\infty} d x x^2 \log \left[1 \mp e^{\left(-\sqrt{x^2+\left(m_i^2(h)+\Pi_i\right) / T^2}\right)}\right],
\end{equation}
where the integral with ``$-/+$ " sign denotes the  contribution of bosons/fermions. $\Pi_i$
is thermal masses of species $i$. Here, we use the daisy resummation scheme proposed by Dolan and Jackiw~\cite{Dolan:1973qd}.
It is worth noticing that only the scalar fields and the longitudinal components of the gauge fields have nonzero $\Pi_i$. 
For the scalar fields
\begin{equation}
	\begin{aligned}
		\Pi_h &= \left(\frac{\lambda_h}{2}+\frac{\lambda_{\phi h}}{12} + \frac{3g^2+g'^2}{16} + \frac{y_t^2}{4}\right)T^2, \quad
		\Pi_\phi &= \frac{\lambda_{\phi h}}{6}T^2\,\,, 
	\end{aligned}
\end{equation}
where $g$ and $g'$ are the gauge coupling of $SU(2)_L$ and $U(1)_Y$, respectively.
For the longitudinal components of the gauge bosons, we have
\begin{equation}
	\Pi_{W_L} = \Pi_{Z_L} = \frac{11g^2}{6}T^2, \quad \Pi_{B_L} =  \frac{11g'^2}{6}T^2\,\,.
\end{equation}
Hence, for the longitudinal components of the gauge bosons, their physical masses are eigenvalues of the following matrix
\begin{equation}
	{M}_{L}^2 = \left(\begin{array}{cccc}
		m_1^2 + \Pi_{W_L} & 0 & 0 & 0\\
		0 & m_1^2 + \Pi_{W_L} & 0 & 0\\
		0 & 0 & m_1^2 + \Pi_{Z_L} & m_{12}^2\\
		0 & 0 & m_{12}^2 & m_2^2 + \Pi_{B_L}\\
	\end{array}\right)\,\,,
\end{equation}
where $m_1^2=g^2 h^2/4$, $m_2^2=g'^2 h^2/4$ and $m_{12}^2=-gg' h^2/4$.

Requiring the ordinary electroweak vacuum with $h=v_0=246~ \mathrm{GeV}$ as the global vacuum at $T=0$ or $V_{\text {eff }}(v_0, 0)<V_{\text {eff }}(0,0)$ leads to
\begin{equation}
\lambda_{\phi h} \lesssim \frac{4 \sqrt{2} \pi m_h}{v_0} \approx 9.0 .
\end{equation}

The phase transition is the process of symmetry breaking in the early Universe. Through a process of bubble nucleation, growth and merger, the Universe transits from a metastable state into a stable state. The critical temperature $T_c$ is defined by the time when the two minima of effective potential are degenerate, $V_{\mathrm{eff}}(v(T_c),T_c)= V_{\mathrm{eff}}(0,T_c)$ with $v(T_c)$ being the vacuum value in the true vacuum at $T = T_c$. Bubbles begin to nucleate when the temperature drops to the nucleation temperature $T_n$.
The nucleation rate of bubbles is given by 
\begin{equation}
	\Gamma(T) \approx T^4\left(\frac{S_3(T)}{2\pi T}\right)^{3/2} e^{-S_3(T) / T}\,\,,
\end{equation}
with $S_3(T)$ being the action of the $\mathcal{O}(3)$ symmetric bounce solution~\cite{Salvio:2016mvj}.
The nucleation temperature $T_n$ is typically defined by
\begin{equation}\label{Tn}
	\Gamma\left(T_n\right) H^{-4}\left(T_n\right) \approx 1 \,\,,
\end{equation}
where $H(T)$ is the Hubble expansion rate,
\begin{equation}
	H^2(T)=\frac{8 \pi}{3 M_{\mathrm{pl}}^2}\left(\frac{\pi^2}{30} g_\star T^4+\Delta V_{\mathrm{eff}}(T)\right)\,\,.
\end{equation}
where $M_{\mathrm{pl}}=1.22 \times 10^{19} \,\,\mathrm{GeV}$  is the Planck mass and $g_\star$ is the number of relativistic degrees of freedom at temperature $T$. $\Delta V_{\mathrm{eff}}(T)$ is the potential energy difference between false and true vacuum
$
	\Delta V_{\mathrm{eff}}(T)=V_{\mathrm{eff}}(0, T)-V_{\mathrm{eff}}(v(T), T)
$.
The potential difference between the inside and outside the bubbles will cause the bubbles to expand in the Universe so that the volume of the false vacuum diminishes with time. The probability of finding a point in the false vacuum reads,
\begin{eqnarray}\label{Tp}
	p(T)=e^{-I(T)}\,\,,
\end{eqnarray}
where $I(T)$ is the fraction of vacuum converted to the true vacuum~\cite{Turner:1992tz,Ellis:2018mja},
\begin{eqnarray}
	I(T)=\frac{4 \pi}{3} \int_T^{T_c} d T^{\prime} \frac{\Gamma\left(T^{\prime}\right)}{T^{\prime 4} H\left(T^{\prime}\right)}\left[\int_T^{T^{\prime}} d \tilde{T} \frac{v_w}{H(\tilde{T})}\right]^3\,\,.
\end{eqnarray}
where $v_w$ is the bubble wall velocity. In the radiation dominated Universe~\cite{Ellis:2018mja},
\begin{equation}
    I(T) = \frac{675M_{\mathrm{pl}}^4v_w^3}{4\pi^5g_\star^2}\int_{T}^{T_c}\frac{dT^{\prime} \Gamma(T^{\prime})}{T^{\prime 6}} \left(\frac{1}{T} - \frac{1}{T^{\prime}}\right)^3\,\,.
\end{equation}
The percolation temperature $T_p$, is defined by $I(T_p) = 0.34$~\cite{Turner:1992tz}. This means that $34\%$ of the false vacuum has been converted to the
true vacuum at $T_p$. The percolation temperature $T_p$ is also the temperature at which the GW is produced  from the FOPT~\cite{Megevand:2016lpr,Kobakhidze:2017mru,Ellis:2020awk,Wang:2020jrd,Ellis:2018mja}.

\begin{figure}[h]
	\begin{minipage}{0.49\linewidth}
		\vspace{3pt}
		\centerline{\includegraphics[width=\textwidth]{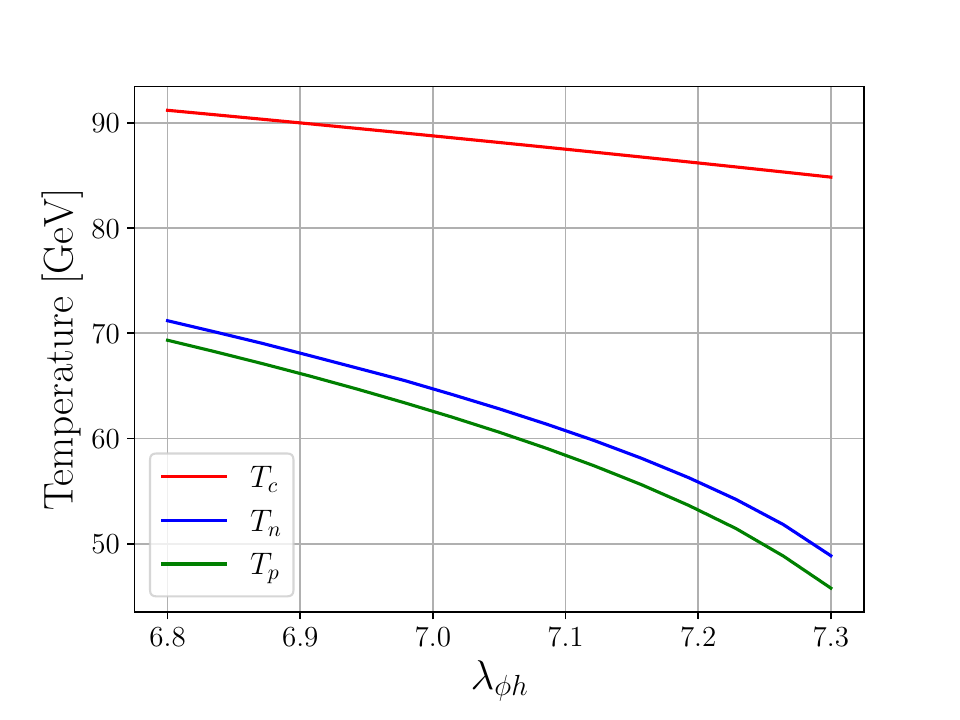}}
	\end{minipage}
	\begin{minipage}{0.49\linewidth}
		\vspace{3pt}
		\centerline{\includegraphics[width=\textwidth]{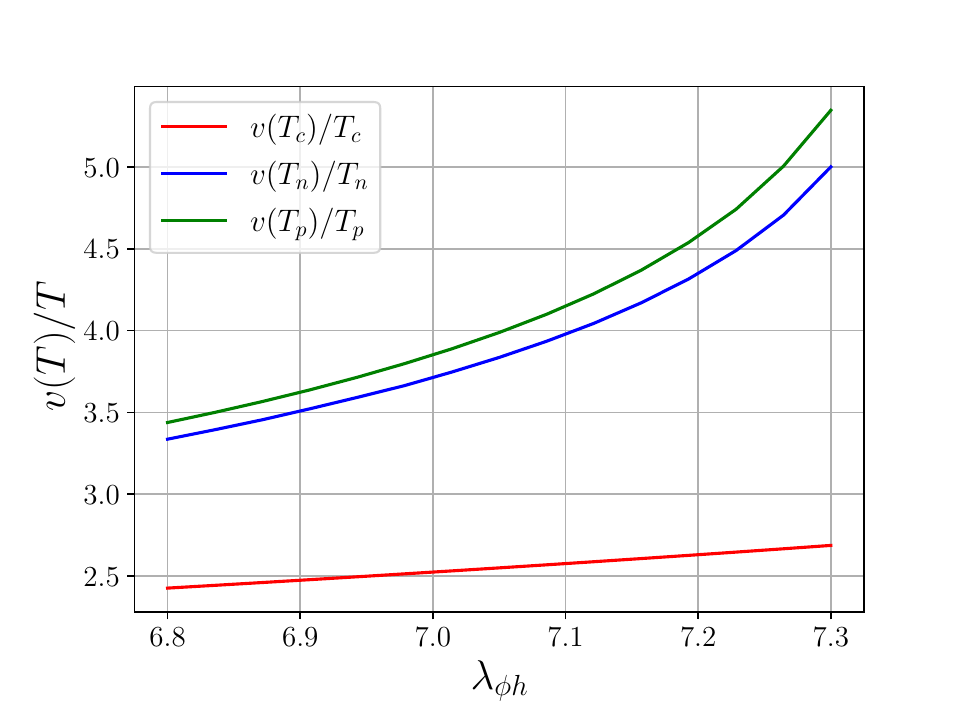}}
	\end{minipage}
	\caption{Phase transition parameters as functions of $\lambda_{\phi h}$. Left: the critical, nucleation and percolation temperatures for different values of $\lambda_{\phi h}$. For the percolation temperature we choose $v_w=0.1$. Right: the wash-out parameters $v(T)/T$ for various temperatures as functions of $\lambda_{\phi h}$.}
	\label{Tphi}
\end{figure}

We use the following definition of the phase transition strength at percolation temperature:
\begin{equation}
\alpha_p \left.\equiv \frac{\left(1-\frac{T}{4} \frac{\partial}{\partial T}\right) \Delta V_{\mathrm{eff}}}{\rho_r}\right|_{T=T_p},
\end{equation}
where $\rho_r=\pi^{2}g_{\star}T^{4}/30$ represents radiation energy density and $\Delta V_{\text {eff }}$ is the potential difference between the false and the true vacua. The inverse time duration $\beta$ at percolation temperature is defined as
\begin{equation}
\frac{\beta}{H_p}= \left.T \frac{d}{d T}\left(\frac{S_3}{T}\right)\right|_{T=T_p}\,\,.
\end{equation}
We use CosmoTransitions~\cite{Wainwright:2011kj} to calculate the phase transition dynamics. In the left panel of figure \ref{Tphi}, we show the three typical temperatures of the FOPT process in the minimal SM plus singlet model. And in the right panel of figure \ref{Tphi}, we show the wash-out parameter $v(T)/T$ at these different temperatures.

\subsection{Bubble wall filtering during FOPT}
As the particles gain mass inside the bubble, due to the energy conservation, only high-energy particles can pass through the bubble walls and the others are reflected. The condition of penetration in the bubble wall frame reads~\cite{Baker:2019ndr,Chway:2019kft}:
\begin{eqnarray}
	p_z^w>\sqrt{\Delta m_i^2}\,\,,
\end{eqnarray}
where $p_z^w$ is the particle $z$-direction momentum in the bubble wall frame, $\Delta m_i^2=(m_i^{\mathrm{in}})^2-(m_i^{\mathrm{out}})^2$ is the mass difference between the true and false vacuum where $m_i^{\mathrm{in}}$ and $m_i^{\mathrm{out}}$ are the particle mass in the true vacuum and false vacuum respectively~\cite{Chao:2020adk}. We will set $m_i^{\mathrm{out}}$ to zero in this work.

The particle flux coming from the false vacuum per unit area and unit time can be written as~\cite{Baker:2019ndr,Chway:2019kft}
\begin{eqnarray}\label{flux}
	{J}_{i}^w=g_i \int \frac{d^3 {p^w}}{(2 \pi)^3} \frac{{p}_z^w}{\sqrt{({{p}^w})^2+(m_i^{\mathrm{out}})^2}} f_i^{\rm eq} \Theta\left({p}_z^w-\sqrt{\Delta m^2}\right) \,\,,
\end{eqnarray}
where $g_{i}$ is the degree of freedom of the particle. $p^w$ is the magnitude of the three-momentum of the particles. $f_i^{\rm eq}$ is the equilibrium distribution outside the bubble in the bubble wall frame,
\begin{eqnarray}\label{feq}
	f_{i}^{\rm eq}=\frac{1}{e^{\gamma_{w}\left(\sqrt{({{p}^w})^2+(m_i^{\mathrm{out}})^2}- v_{w} {p}_z^w\right) / T}\mp 1}\,\,,
\end{eqnarray}
where $\mp$ is for bosons and fermions respectively.  $\gamma_w=1/\sqrt{1-v_w^2}$ is the Lorentz boost factor. 
The particle number density inside the bubble $n_{i}^{\text {in }}$ in the bubble center frame can be written as 
\begin{eqnarray}\label{nin}
	n_{i}^{\text {in }}=\frac{{J}_{i}^w}{  \gamma_{w}   v_{w}}\,\,.
\end{eqnarray}
Assuming the particles are massless in the false vacuum, 
 we can integrate Eq.~\eqref{flux} analytically and get~\cite{Chway:2019kft,Jiang:2023nkj}
\begin{eqnarray}\label{nana1}
	n_i^{\text {in }} \simeq \frac{g_{i} T_{}^3}{\gamma_w v_w}\left(\frac{\gamma_{w}\left(1- v_{w}\right) m_i^{\mathrm{in}} / T_{}+1}{4 \pi^2  \gamma_{w}^3\left(1- v_{w}\right)^2}\right) e^{-\frac{ \gamma_{w}\left(1-  v_{w}\right) m_i^{\mathrm{in}}}{T_{}}} \,\,,
\end{eqnarray}
where we have used Maxwell-Boltzmann approximation of DM distribution.
One can see that as ${v}_{w} \rightarrow 1$,  Eq.~\eqref{nana1} approaches $n_i^{\mathrm{out}}=g_i T^3 / \pi^2$, which is approximately the equilibrium number density for Boltzmann distribution outside the bubble. 
The fraction of particles $i$ that are trapped into the false vacuum is defined by
\begin{equation}
	F_i^{\mathrm{trap}} = 1-\frac{n_i^{\mathrm{in}}}{n_i^{\mathrm{out}}}\,\,.
\end{equation}
In our case, complex scalars $\phi$ and $\phi^{\dagger}$ are trapped into the false vacuum due to the filtering effect. The symmetric part would annihilate away in terms of the process $\phi+\phi^{\dagger}\rightarrow h+h$, then only the asymmetric part survives and composes the charge of gauged Q-balls. 
It can be easily seen that the penetrated particle number density is sensitive to the bubble wall velocity as it appears in the exponent of Eq.~\eqref{nana1}. The precise calculation of bubble wall velocity~\cite{Moore:1995si,Laurent:2022jrs,Lewicki:2021pgr,Wang:2020zlf,Jiang:2022btc,Dorsch:2021nje,DeCurtis:2022hlx,Ai:2023see} is beyond the scope of this work and we set it as a free parameter.

\subsection{Charge of Q-ball in the electroweak FOPT}

We can
define $T_{\star}$ as the temperature at which the false vacuum or old phase remnants can still form an 
infinitely connected ``cluster", just like the definition of percolation temperature~\cite{Hong:2020est}. The $T_\star$ satisfies $p(T_\star)=1-p(T_p)=0.29$ which corresponds to $I(T_\star)=1.24$.
$T_\star$ is the temperature when Q-balls start to form. Below the temperature $T_\star$, the false vacuum remnants formed during FOPT may further fragment into smaller pieces. Ultimately, these pieces would shrink into Q-balls if there exists a non-zero primordial charge asymmetry. We can define the critical radius,  $r_c$, at which the remnant shrinks to an insignificant size before another true vacuum bubble forms within it~\cite{Krylov:2013qe}. This means~\cite{Hong:2020est}
\begin{equation}
	\Gamma\left(T_\star\right) \left(\frac{4 \pi}{3} r_{c}^3\right) \Delta t \sim 1\,\,.
\end{equation}
where $\Delta t=r_{c}/v_{w}$ is the time cost for shrinking. The number density of the remnants  $n_{\mathrm{Q}}^{\star}$ can be expressed as: 
\begin{equation}
	n_{\mathrm{Q}}^{\star}\simeq 0.29 \left(\frac{3}{4 \pi}\right)^{1 / 4}\left(\frac{\Gamma\left(T_\star\right)}{v_w}\right)^{3 / 4} \,\,,
\end{equation}
since the condition $n_{\mathrm{Q}}^\star \left(\frac{4 \pi}{3} r_{c}^3\right)=p(T_\star)\simeq 0.29$. 

The formation of Q-balls requires a nonzero conserved primordial charge which comes from the primordial DM asymmetry $\eta_{\phi}=(n_{\phi}-n_{\phi^{\dagger}})/s$ with entropy density $s_{}=2\pi^{2}g_{\star}T_{}^{3}/45$. 
If the DM asymmetry is produced by thermal freeze-out in the early Universe, it is bounded from above by the equilibrium value,
\begin{equation}
\eta_\phi \lesssim \eta_\phi^{\mathrm{eq}} =  \frac{n_\phi^{\mathrm{eq}}(T)}{s(T)} \simeq 5.1 \times 10^{-3} \times \left(\frac{108.75}{g_\star}\right)\,\,,
\end{equation}
where we have used $n_\phi^{\mathrm{eq}}(T) = 2\zeta(3) T^3/\pi^2$ with  $\zeta (3) =1.20206$ being the value of Riemann zeta function $\zeta(s)$ at $s=3$. In order to overcome this constraint, we assume the DM is produced by some non-thermal processes like decay. In this work, we do not specify the origin of primordial charge of the complex scalar $\phi$.
In the early Universe at higher temperature, new physical processes beyond the standard model may have occurred.  The would-be Q-ball DM particles $\phi$
may have their own conserved charge and be created in asymmetric
decays of heavier particles~\cite{Kaplan:2009ag}. For example, heavy Majorana neutrino could decay into a light fermion and a scalar, like $N \rightarrow \bar{\chi} + \phi$ and $N \rightarrow \chi + \phi^{\dagger}$ where $\chi$ is a fermion~\cite{Falkowski:2011xh}. Assuming the process is CP-violating, the decay rates of these two channels differentiate from each other at loop level (this is similar to the process of leptogenesis.). So the asymmetry between $\phi$ and $\phi^{\dagger}$ appears and is retained until the phase transition in this work. The large DM asymmetry can be discussed in a similar manner to the large lepton asymmetry in the leptogenesis scenarios. Recent  measurement of ${}^4\mathrm{He}$  abundance coming from the EMPRESS experiment suggests a large degeneracy parameter of the electron neutrino~\cite{Matsumoto:2022tlr},
\begin{equation}
\xi_e=0.05_{-0.02}^{+0.03}\,\,.
\end{equation}
Since the neutrino oscillations among three flavors, the neutrinos with three flavors have the same amount of asymmetry, and then the total lepton asymmetry reads,
\begin{equation}
    \eta_L \equiv 3\frac{n_{\nu_e}-n_{\bar{\nu}_e}}{s} = 3\frac{T^3\xi_e/6}{2\pi^2g_{BBN}T^3/45} \simeq 5.3 \times 10^{-3}\,\,,
\end{equation}
where $g_{BBN}=10.75$ is the relativistic degree of freedom at the epoch of big bang nucleosynthesis. The large lepton asymmetry may come from the low-scale leptogenesis~\cite{Borah:2022uos,ChoeJo:2023cnx}, Affleck-Dine mechanism~\cite{McDonald:1999in} or L-ball decay~\cite{Kawasaki:2022hvx}.

\begin{figure}[h]
\centering
	\begin{minipage}{0.78\linewidth}
		\vspace{3pt}
		\centerline{\includegraphics[width=\textwidth]{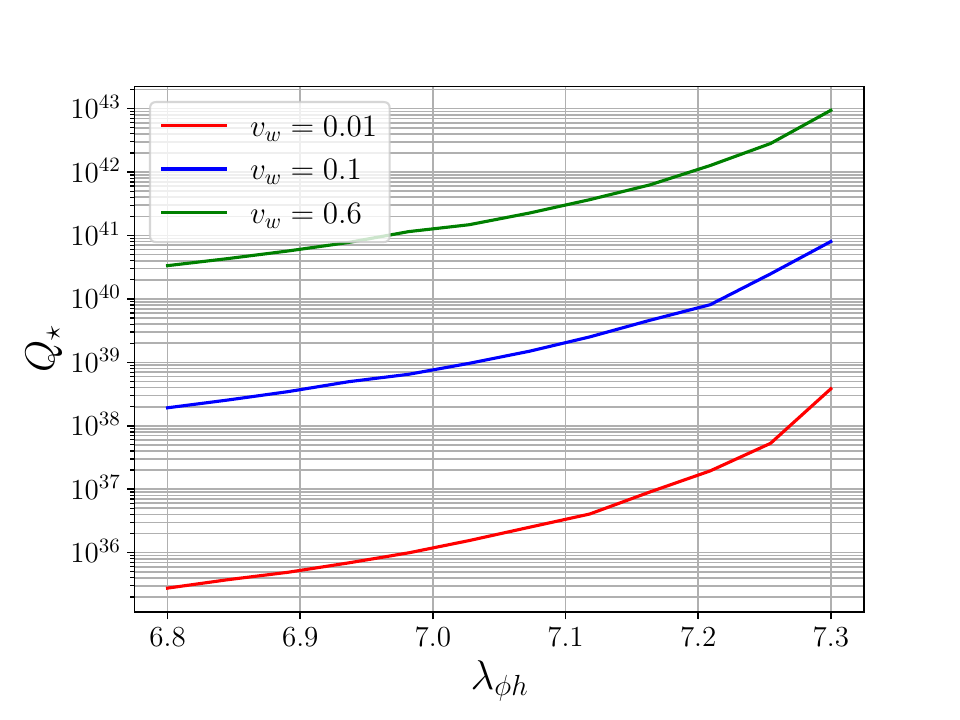}}
	\end{minipage}
	\caption{Charge of gauged Q-ball in electroweak FOPT. Here we choose $\eta_\phi = \eta_L$.}
	\label{Qstar}
\end{figure}

In a remnant, the trapped Q-charge is given by 
$Q_{\star}=F_{\phi}^{\mathrm{trap}} \eta_\phi s_{\star}/{n_{\mathrm{Q}}^{\star}}$. In figure \ref{Qstar}, we show the charge of the gauged Q-ball DM formed during the FOPT for different values of bubble wall velocities. We have chosen $\eta_\phi=\eta_L$.  When $\lambda_{\phi h}$ is larger, both 
the $\Gamma(T_\star)$ and the Q-ball number density are suppressed, so that the charge is larger at a given $\eta_\phi$. 

Since $n_{\mathrm{Q}} / s$ and $Q$ does not change in the adiabatic Universe, at present they are
\begin{equation}
	n_{\mathrm{Q}}=\frac{n_{\mathrm{Q}}^{\star}}{s_\star} s_0, \quad Q=Q_{\star}\,\,,
\end{equation}
with $s_0=2891.2~\mathrm{cm}^{-3}$ being the entropy density in current time~\cite{ParticleDataGroup:2018ovx}.

We take the approximation $T_\star \approx T_p$ then the bounce action can be approximated written as~\cite{Huber:2007vva}
\begin{equation}\label{S3T}
	\frac{S_3(T_\star)}{T_\star} \simeq 131 - 4\ln \left(\frac{T_\star}{100~\mathrm{GeV}}\right) - 4\ln \left(\frac{\beta/H_\star}{100}\right) + 3\ln v_w -2\ln \left(\frac{g_\star}{100}\right)\,\,,
\end{equation}
where $\beta/H_\star=\left. \beta/H \right|_{T=T_\star}$. By approximately using $\Gamma(T_\star) \approx T_\star^4 e^{-S_3(T_\star)/T_\star}$, we get
\begin{equation}
	Q_{} = 5.35\times 10^{36} F_\phi^{\mathrm{trap}} \left(\frac{v_w}{0.01}\right)^3\left(\frac{\eta_\phi}{\eta_L}\right)\left(\frac{100~\mathrm{GeV}}{T_\star}\right)^3\left(\frac{100}{\beta/H_\star}\right)^3\left(\frac{100}{g_\star}\right)^{1/2}\,\,.
\end{equation}
In reality, the mass or charge of Q-balls should have a distribution, 
which has been discussed in detail in Ref.~\cite{Lu:2022paj}. The false vacuum size distribution is given by Eq.~(15) of Ref.~\cite{Lu:2022paj}.
However, it has been shown in Eq.~(24) of Ref.~\cite{Lu:2022paj} that the average size of Q-balls during a FOPT is still approximately equal to $v_w/\beta$, just as in the monochromatic case. Based on this, for simplicity we can still use the average value instead of the distribution of false vacuum. In order to get stable gauged Q-ball with a given value of charge, we must impose $Q_\mathrm{max}>Q_{}$, so the gauge coupling $\tilde{g}$ has to satisfy
\begin{equation}\label{gmax}
	\tilde{g}<1.28\times 10^{-18} \left(\frac{1}{F_\phi^{\mathrm{trap}}}\right)^{1/2}\left(\frac{0.01}{v_w}\right)^{3/2}\left(\frac{\eta_L}{\eta_\phi}\right)^{1/2}\left(\frac{T_\star}{100~\mathrm{GeV}}\right)^{3/2}\left(\frac{\beta/H_\star}{100}\right)^{3/2}\left(\frac{g_\star}{100}\right)^{1/4}\,\,.
\end{equation}

\subsection{Relic density of gauged Q-ball DM}

The DM relic density can also receive the contribution from the standard freeze-out process, through the process $\phi +\phi^{\dagger}\leftrightarrow h+h$. The relic abundance reads,
\begin{equation}
	\Omega_{\mathrm{freeze-out}}h_{100}^2 \approx \frac{2.58\times 10^{-10}~\mathrm{GeV}^{-2}}{\langle \sigma_{\mathrm{anni}} v_{\mathrm{rel}}\rangle}\,\,,
\end{equation}
where $\langle \sigma_{\mathrm{anni}} v_{\mathrm{rel}}\rangle$ is the annihilation cross section. $h_{100}=H_0 /\left(100 \mathrm{~km} \cdot \mathrm{s}^{-1} \cdot \mathrm{Mpc}^{-1}\right)=0.67$ where $H_0$ is the Hubble constant today~\cite{Planck:2018vyg}. The cross section of process $\phi +\phi^{\dagger}\leftrightarrow h+h$ reads $\frac{\lambda_{\phi h}^2}{64\pi m_\phi^2}\left(1-\frac{m_h^2}{m_\phi^2}\right)^{1/2}$~\cite{McDonald:1993ex}. In our parameter space, where the $\lambda_{\phi h}$ is around 7, the relic abundance from freeze-out is approximately $\Omega_{\mathrm{freeze-out}}h_{100}^2 \approx 2.35\times 10^{-4}$.
So we can omit the DM produced from thermal freeze-out.

The DM relic density also receives the contribution from penetrated asymmetric components of DM particles which is given by the excess of $\phi$ over $\phi^{\dagger}$,
\begin{equation}\label{asymmetric}
\Omega_{\mathrm{asymmetric}}h_{100}^2=(1-F_{\phi}^{\mathrm{trap}})\eta_{\phi}s_0 m_\phi\,\,.
\end{equation}

The relic density of Q-balls at present is
\begin{equation}
\Omega_{\mathrm{Q}} h_{100}^2=\frac{n_{\mathrm{Q}} E_{\mathrm{Q}}}{\rho_{\mathrm{c}}} h_{100}^2\,\,,
\end{equation}
where $\rho_{\mathrm{c}}=3 H_0^2 M_{\mathrm{pl}}^2 /(8 \pi)$ is the critical energy density. We have found that, the gauged Q-ball is generally a mixed state of Eq.~\eqref{abrm1} and Eq.~\eqref{abrl1} as $0\leq \alpha B_0 \rho_b \leq C_2$. So we can write down the energy of gauged Q-ball at a given charge, 
\begin{equation}
	E_Q \simeq \frac{\pi Q}{r_\star} + \frac{4\pi }{3}r_\star^3V_0 + \frac{3}{20\pi}\frac{\tilde{g}^2Q^2}{r_\star}\,\,,
\end{equation}
where $V_0=\frac{\lambda_h}{4}v_0^4$ is the potential difference between inside and outside of the gauged Q-balls at zero temperature. The first term on the right-side is the zero-point energy of the scalar particles, the second term is the vacuum volume energy inside the Q-ball and the third term is the  electrostatic self energy. By minimizing this expression respect to $r_\star$, we obtain,
\begin{equation}
    E_Q = \frac{4\pi }{3} \left(4V_0\right)^{1/4}Q^{3/4}\left(1+\frac{3\tilde{g}^2 Q}{20\pi^2}\right)^{3/4}\,\,.
\end{equation}
In the limit of zero gauge coupling,
\begin{equation}
	E_Q =\frac{4\pi }{3} \left(4V_0\right)^{1/4}Q^{3/4}=\frac{4\pi }{3}Q^{3/4}\lambda_h^{1/4}v_0\,\,,
\end{equation}
which is just the energy of global Q-ball. By using $n_{\mathrm{Q}}=n_{\mathrm{Q}}^{\star}s_0/s_\star$ and $Q=Q_{\star}=F_{\phi}^{\mathrm{trap}} \eta_\phi s_{\star}/{n_{\mathrm{Q}}^{\star}}$, we finally arrive at the expression:
\begin{equation}
\begin{aligned}
	&\Omega_{\mathrm{Q}} h_{100}^2 \\
 &\simeq 2.81 \times \left(\frac{s_0h_{100}^2}{\rho_c}\right) \left(\frac{\Gamma(T_\star)}{v_w}\right)^{3/16}s_\star^{-1/4}(F_\phi^{\mathrm{trap}}\eta_\phi)^{3/4} \lambda_h^{1/4}v_0\left(1+\frac{108^{1/4}\tilde{g}^2F_{\phi}^{\mathrm{trap}}\eta_{\phi}s_\star v_w^{3/4}}{5.4\pi^{7/4}\Gamma(T_\star)^{3/4}}\right)\,\,,
\end{aligned}
\end{equation}
the $\Gamma(T_\star)$ can also be expanded by using $\Gamma(T_\star) \approx T_\star^4 e^{-S_3(T_\star)/T_\star}$ and Eq.~\eqref{S3T}, but we keep the expression here to give more accurate results.

Although the expression hitherto is general, we apply these to the minimal SM plus singlet model. We choose $\lambda_{\phi h}=6.8$ for which $T_n=71.65~\mathrm{GeV}$ and $T_p= 68.9~\mathrm{GeV}$. The value of this portal coupling satisfies the validity of the perturbative analysis which indicates the portal coupling should be roughly smaller than 10~\cite{Curtin:2014jma}. 
The number density of gauged Q-balls at production is $n_Q^{\star} \simeq 3.0\times 10^{-31}~\mathrm{GeV}^{-3}$. In this case, the value of $v(T_p)/T_p$ is approximately 3.5 and there are still 50\% of the DM particles trapping inside the false vacuum even at $v_w=0.6$. However, it should be noted that in this case, the contribution from penetrated asymmetric DM will dominate, as can be seen in Eq.~\eqref{asymmetric}. This can be avoided in two ways. One way is to increase the phase transition strength and the corresponding $v(T)/T$ so there are little DM particles penetrating into the true vacuum, 
resulting $F_\phi^{\mathrm{trap}}\approx 1$. This can be achieved by introducing new freedoms beyond the standard model or considering dark FOPT instead of electroweak FOPT. The other way is model dependent: 
one can introduce new decay channels that penetrated $\phi$ could decay into 
dark radiation or SM leptons which can account for the lepton asymmetry. The decay process does not destroy the stability of Q-balls as long as $\pi/r_\star < m_d$ where $m_d$ is the mass of decay products. In this work, we focus on the gauged Q-ball DM so we do not consider the penetrated asymmetric DM in detail.


\begin{figure}[h]
\centering
	\begin{minipage}{0.78\linewidth}
		\vspace{3pt}
		\centerline{\includegraphics[width=\textwidth]{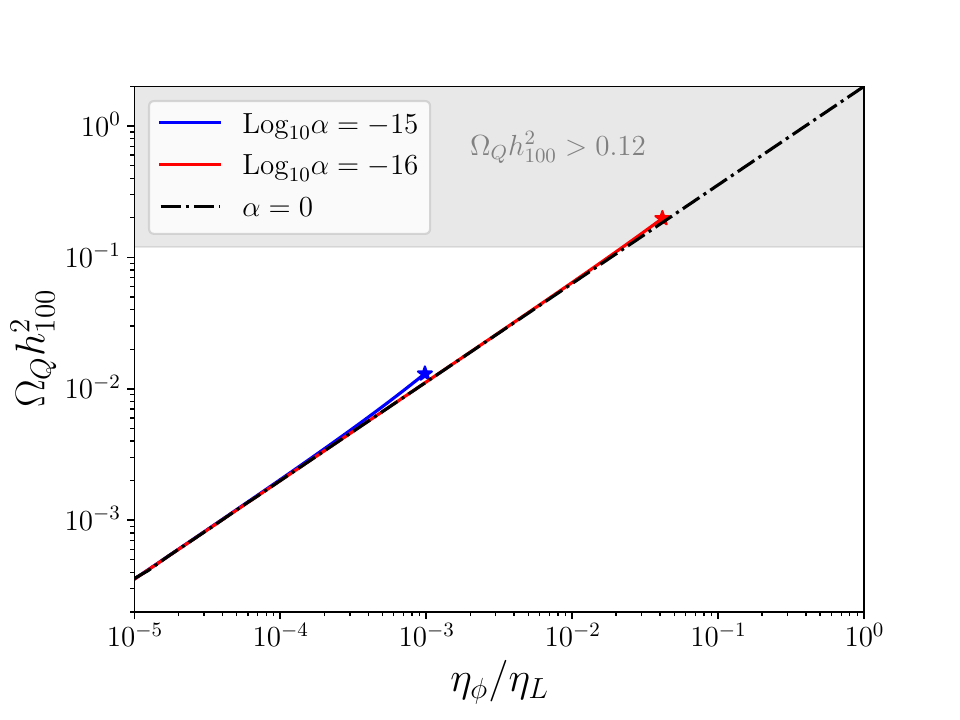}}
	\end{minipage}
	\caption{ Q-ball relic density as function of DM asymmetry $\eta_\phi/\eta_L$ in the SM plus singlet model where $\lambda_{\phi h}=6.8$. We choose $v_w=0.01$. The star represents the value for the gauged Q-ball with maximal charge. The gray region represents the region where the DM is overproduced.}
	\label{Omegah2}
\end{figure}

We show the gauged Q-ball DM relic density in figure \ref{Omegah2}. The colored stars represent the values for the gauged Q-ball with maximal charge. The gray region represents the region where the DM is overproduced. We can see that due to the finiteness of the charge, the gauged Q-balls can explain the whole DM at $v_w=0.01$ only when the rescaled gauge coupling $\alpha \lesssim 10^{-16}$. The DM relic density is slightly enhanced due to the extra electrostatic energy.

In table~\ref{ptable} we choose four benchmark points that satisfy the correct DM relic density and show the corresponding $F_{\phi}^{\mathrm{trap}}$ and $\eta_{\phi}/\eta_L$. The condition of a strong FOPT leads to obvious deviation of triple Higgs coupling, which might be detected by the loop-induced $e^+ e^-\to hZ$ process~\cite{Huang:2015izx, Huang:2016odd} at future lepton colliders, such as FCC-ee, CEPC, and ILC. 
One can define the $\delta \sigma_{Zh}$ as the fractional change in $Zh$ production relative to the SM prediction at one loop. We list the corresponding $\delta \sigma_{Zh}$ for the four benchmark points in table~\ref{ptable}.
We can see that the value of the required DM asymmetry $\eta_\phi$ is close to the lepton asymmetry $\eta_L$. This prompts us to speculate that they have the same origin. Actually, if the DM asymmetry comes from the process $N \rightarrow \bar{\chi} + \phi$ and $N \rightarrow \chi + \phi^{\dagger}$. We could assume the dark fermion $\chi$ is long-lived and decay suddenly into leptons after electroweak phase transition in order to avoid the electroweak sphaleron process. Then we expect the $\eta_\phi$ and $\eta_L$ is at the same order. The detailed model building is beyond the scope of this work and we leave this in our future studies.

\begin{table}[t]
	\centering
	
	\setlength{\tabcolsep}{3mm}
	
	\begin{tabular}{c|c|c|c|c|c|c|c|c|c}
		\hline\hline
		& $\lambda_{\phi h}$ & $T_p$~[GeV] &$\alpha_p$ &$\beta/H_p$ &$v_w$ & $F_{\phi}^{\mathrm{trap}}$  & $\eta_{\phi}/\eta_L$ &  $\delta \sigma_{Zh}$& GW\\
		\hline
        $BP_1$	 &6.8 & 69.8 & 0.12 &540  &$0.1$ & 0.932&0.48  & -0.36\% &$\textcolor{cyan}{\bullet}$\\
            \hline
		$BP_2$	 &6.8 & 70.4 & 0.12 &578  &$0.6$ & 0.805&3.0  & -0.36\% &$\textcolor{magenta}{\bullet}$\\
            \hline
		$BP_3$	&7.0 & 63.0 & 0.15 & 372 &$0.1$ & 0.965 &3.4& -0.37\% &$\textcolor{black}{\bullet}$\\
            \hline
		$BP_4$	&7.0 & 63.9 & 0.15 & 403  &$0.6$ & 0.858 & 20.8  &-0.37\%  &$\textcolor{red}{\bullet}$\\
		\hline\hline
	\end{tabular}
	\caption{Parameters of the model and phase transitions that satisfy $\Omega_{\mathrm{Q}} h_{100}^2\equiv 0.12$. We choose $\alpha=\tilde{g}/\sqrt{2\lambda_h}=10^{-18}$. $F_{\phi}^{\mathrm{trap}}$ is the fraction of DM particles which are trapped inside the false vacuum and $\eta_\phi/\eta_L$ is the DM asymmetry compared with the lepton asymmetry. $\delta \sigma_{Zh}$ is the fractional change in $Zh$ production relative to the SM prediction at one loop. GW represents the corresponding GW spectra in figure \ref{gw}.}\label{ptable}
\end{table}

\section{Constraints and detection of gauged Q-ball DM}\label{signalconstraints}
\subsection{Direct detection and astronomical constraints}
\begin{figure}[h]
\centering
	\begin{minipage}{0.8\linewidth}
		\vspace{3pt}
		\centerline{\includegraphics[width=\textwidth]{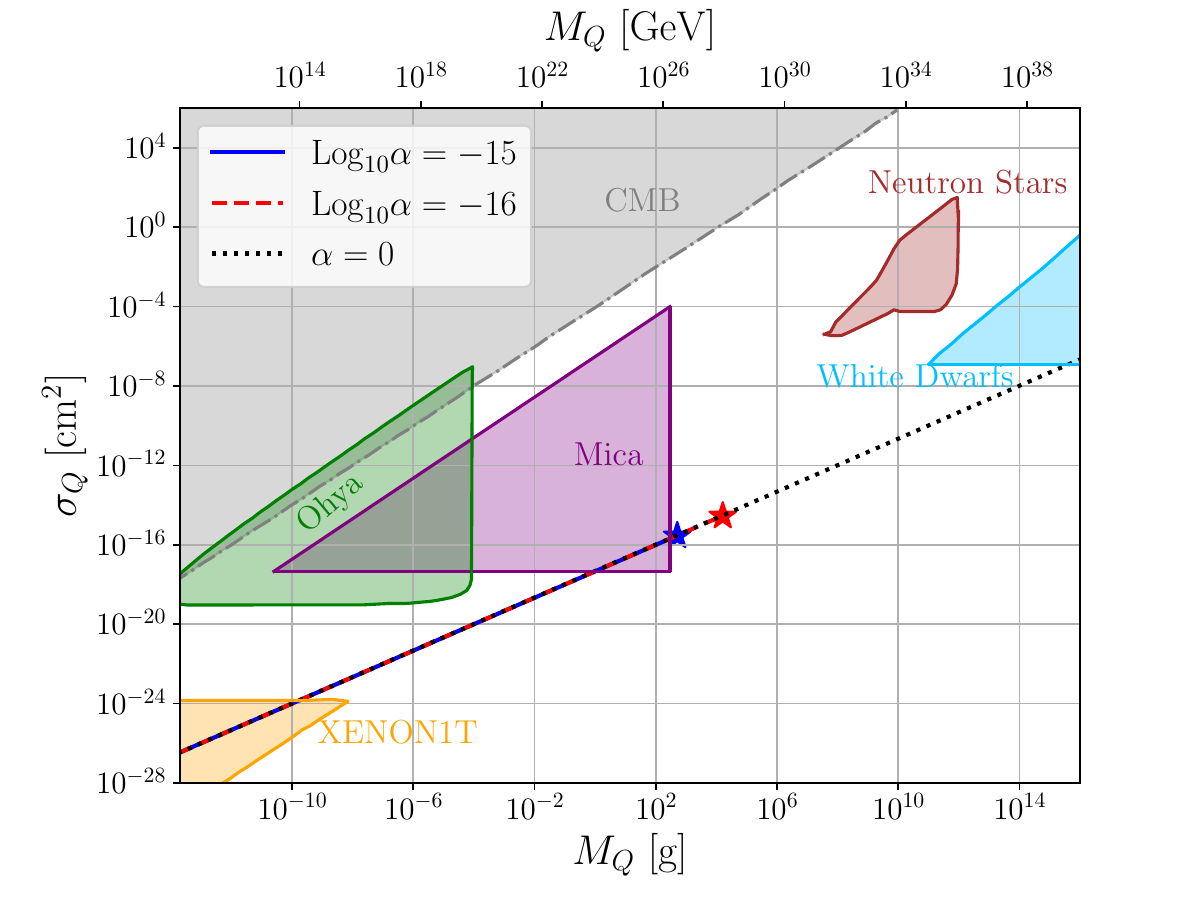}}
	\end{minipage}
	\caption{Combined constraints of direct detection and astronomical experiments on gauged Q-ball DM in the SM plus singlet model where $\lambda_{\phi h}=6.8$. The value corresponding to maximal charge is marked by a star.}
	\label{constraints}
\end{figure}

As we have mentioned, the mass/charge distribution of gauged Q-balls formed during FOPT is not
monochromatic. We keep the gauged Q-ball DM mass and radius as free parameters to discuss its detection potential or constraints. The combined constraints of gauged Q-ball is shown in figure \ref{constraints}. As we have discussed in the previous sections, the size of the gauged Q-ball is restricted to be finite. The maximal charge or the maximal mass of gauged Q-ball DM with given values of gauge coupling are marked by the stars in figure \ref{constraints}.
The gray region denotes the constraints from cosmic microwave
background (CMB) which is affected by DM-baryon scattering~\cite{Dvorkin:2013cea,Jacobs:2014yca}. DM cross sections have been probed by a variety of shallow and deep
underground DM or repurposed experiments: XENON1T~(orange)~\cite{XENON:2018voc,Clark:2020mna}, Mica~(purple)~\cite{Price:1986ky,Price:1988ge}, Ohya~(green)~\cite{Bhoonah:2020fys}. The Q-ball DM would transfer
energy with other objects primarily through elastic scattering, approximately utilizing their geometric cross-section. When these cross-sections are sufficiently large, the resultant linear energy transfer might trigger observable phenomena. For example, if a
macroscopic DM traverses compact objects such as white dwarfs or neutron stars and triggers
thermonuclear runaway, this could potentially lead to a type IA supernova or superburst, respectively~\cite{Graham:2018efk}. The brown region represents constraints from superbursts in neutron stars~\cite{SinghSidhu:2019tbr} and the blue region from white dwarf becoming supernovae~\cite{Graham:2018efk,SinghSidhu:2019tbr}. Combined all constraints in figure \ref{constraints}, one can see that the gauged Q-balls could be the DM candidate in a wide region of parameter space.

\subsection{Phase transition GW}

The phase transition GW spectra come from three sources of a strong FOPT: bubble collision, sound wave, and turbulence.
\begin{itemize}
	\item Bubble collision\\
	The formula of phase transition GW from bubble collisions~\cite{Huber:2008hg,Caprini:2015zlo} at the percolation temperature $T_{p}$~, reads:
	\begin{equation}
		 \Omega_{\mathrm{co}}h_{100}^2 \simeq 1.67 \times 10^{-5}\left(\frac{H_p}{\beta}\right)^2\left(\frac{\kappa_h \alpha_p}{1+\alpha_p}\right)^2\left(\frac{100}{g_{\star}}\right)^{1 / 3} \frac{0.11 v_w^{3}}{0.42+v_w^2} \frac{3.8\left(f / f_{\mathrm{co}}\right)^{2.8}}{1+2.8\left(f / f_{\mathrm{co}}\right)^{3.8}}\,\,,
	\end{equation}
	where $ \kappa_h $ represents the fraction of vacuum energy converted into the scalar field's gradient energy. $f_{\mathrm{co}}$ is the peak frequency of bubble collision:
	\begin{equation}
		f_{\mathrm{co}} \simeq 1.65 \times 10^{-5} \mathrm{~Hz} \left(\frac{\beta}{H_p }\right)\left(\frac{0.62}{1.8-0.1 v_w+v_w^2}\right)\left(\frac{T_{p}}{100~ \mathrm{GeV}}\right)\left(\frac{g_\star}{100}\right)^{1 / 6}\,\,.
	\end{equation}
	\item Sound wave\\
The contribution from sound waves could be more significant. The formula GW spectrum from sound waves is~\cite{Hindmarsh:2017gnf}:
	\begin{equation}
		\begin{aligned}
		\Omega_{\mathrm{sw}}h_{100}^2 \simeq & 2.65 \times 10^{-6} \Upsilon_{\mathrm{sw}} \left(\frac{H_p}{\beta}\right)\left(\frac{\kappa_v \alpha_p}{1+\alpha_p}\right)^2\left(\frac{100}{g_{\star}}\right)^{1 / 3} 
			  v_w  \left(f / f_{\mathrm{sw}}\right)^3  \left(\frac{7}{4+3\left(f / f_{\mathrm{sw}}\right)^2}\right)^{7 / 2}\,\,,
		\end{aligned}
	\end{equation}
where $  \kappa_v $ reflects the fraction of vacuum energy that transfers into the fluid's bulk motion. The peak frequency of sound wave processes is:
\begin{equation}
		f_{\mathrm{sw}} \simeq 1.9 \times 10^{-5} \mathrm{~Hz} \frac{1}{v_w}\left(\frac{\beta}{H_p}\right)\left(\frac{T_{p}}{100~ \mathrm{GeV}}\right)\left(\frac{g_\star}{100}\right)^{1 / 6}\,\,.
\end{equation}
$\Upsilon_{\mathrm{sw}}$ is the suppression factor of the short period of the sound wave,
\begin{equation}
\Upsilon_{\mathrm{sw}} = \left(1-\frac{1}{\sqrt{1+2\tau_{\mathrm{sw}}H_p}}\right)\,\,,
\end{equation}
where
\begin{equation}
    \tau_{\mathrm{sw}}H_p \approx (8\pi)^{\frac{1}{3}}\frac{v_w (H_p/\beta)}{ \sqrt{3\kappa_v\alpha_p/(4+4\alpha_p)}}\,\,.
\end{equation}
\item Turbulence  \\
The formula of the GW spectrum from turbulence is~\cite{Binetruy:2012ze}:
\begin{equation}
	 \Omega_{\mathrm{turb}}h_{100}^2 \simeq 3.35 \times 10^{-4} \left(\frac{H_p v_w}{\beta}\right)\left(\frac{\kappa_{\mathrm{turb}} \alpha_p}{1+\alpha_p}\right)^{3 / 2}\left(\frac{100}{g_\star}\right)^{1 / 3} \frac{\left (f / f_{\text {turb }}\right)^3}{\left(1+f / f_{\text {turb }}\right)^{11 / 3}\left(1+8 \pi f / H_p\right)}\,\,,
\end{equation}	
note that $H_p$ is the Hubble rate at $T_p$:
 \begin{equation}
H_p=1.65 \times 10^{-5} 
~\mathrm{Hz}\left(\frac{T_{p}}{100~\mathrm{GeV}}\right)\left(\frac{g_\star}{100}\right)^{1/6}\,\,,
\end{equation}
and the peak frequency of turbulence processes is :
	\begin{equation}
		f_{\text {turb }} \simeq 2.7 \times 10^{-5} \mathrm{~Hz}\frac{1}{v_w} \left(\frac{\beta}{H_p}\right)\left(\frac{T_{p}}{100~ \mathrm{GeV}}\right)\left(\frac{g_\star}{100}\right)^{1 / 6}\,\,.
  \end{equation}
$ \kappa_{\mathrm{turb} }$ represents the efficiency of vacuum energy being converted into turbulent flow:
\begin{equation}
		\kappa_{\text {turb }}=\tilde{\epsilon} \kappa_v\,\,,
\end{equation}
where the $\tilde{\epsilon}$ is set to $0.1$.
\end{itemize}

The total contribution to the GW spectra can be calculated by summing these individual contributions:
\begin{equation}
\Omega_{\mathrm{GW}}h_{100}^2=\Omega_{\mathrm{co}}h_{100}^2+ \Omega_{\mathrm{sw}}h_{100}^2+ \Omega_{\mathrm{turb}}h_{100}^2\,\,.
\end{equation}

\begin{figure}[h]
\centering
	\begin{minipage}{0.7\linewidth}
		\vspace{3pt}
		\centerline{\includegraphics[width=\textwidth]{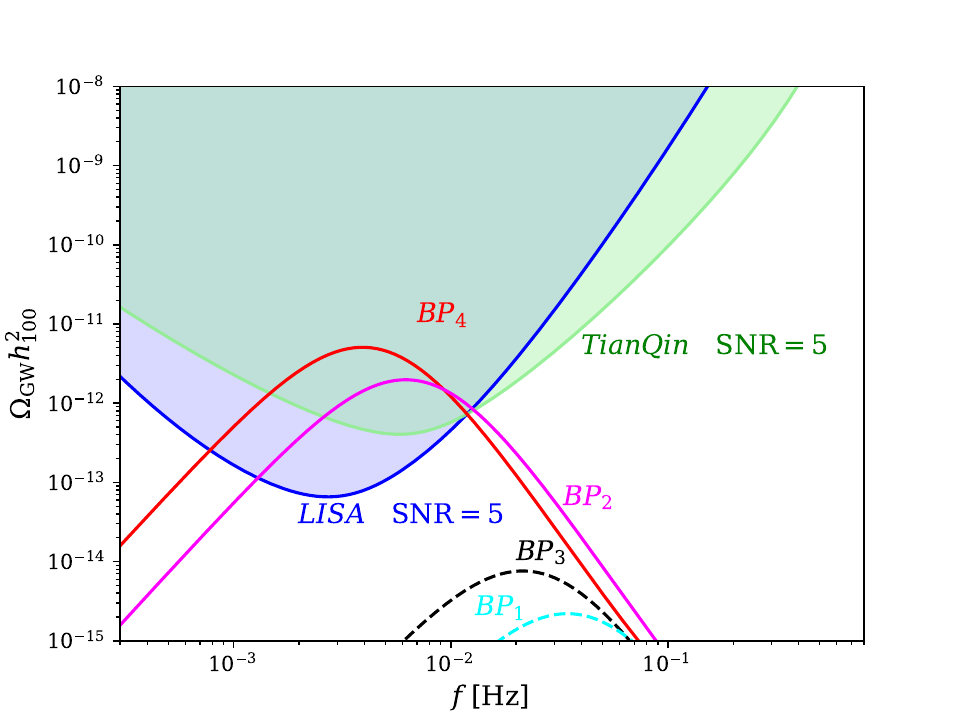}}
	\end{minipage}
	\caption{GW spectra for four benchmark points in  Table \ref{ptable}. The future GW detectors LISA~\cite{LISA:2017pwj},
TianQin~\cite{TianQin:2015yph,Liang:2022ufy} could detect $BP_2$ and $BP_4$ whose bubble wall velocities are relatively large.}
	\label{gw}
\end{figure}

We show the GW spectra $\Omega_{\mathrm{GW}}h_{100}^2$ for four benchmark points in figure \ref{gw}. These four benchmark points are shown in Table~\ref{ptable}. We choose $\lambda_{\phi h}=6.8$ and $7.0$ and for each value of $\lambda_{\phi h}$ we choose two bubble wall velocities $v_w=0.1,0.6$. The colored regions represent the sensitivity
curves for future GW detectors LISA~\cite{LISA:2017pwj} and TianQin~\cite{TianQin:2015yph,Liang:2022ufy} with the signal-to-noise ratio (SNR) about 5. We can see that the LISA and
TianQin could detect this new DM mechanism when the bubble wall velocity is relatively large. Taiji~\cite{Hu:2017mde}, BBO~\cite{Corbin:2005ny}, DECIGO~\cite{Seto:2001qf}, Ultimate-DECIGO~\cite{Kudoh:2005as} could also detect this new DM mechanism by GW signals.

\section{Discussion and conclusion}\label{conclusion}

In this work, we have systematically discussed the gauged Q-ball DM formed 
during a FOPT. We have investigated the stability of gauged Q-balls, including quantum stability, stress stability, fission stability, and classical stability. Different from the global Q-balls, the gauge interaction restricts the size of the stable gauged Q-balls. For a given value of the gauge coupling, the stable
gauged Q-balls can only be realized in the region of charge $Q_s<Q<Q_{\mathrm{max}}$. The upper limit $Q_{\mathrm{max}}$ and lower limit $Q_s$ mainly come from the stress stability and quantum stability criterion respectively. By using the thin-wall approximation, we show that the piecewise model can describe the basic properties of gauged Q-balls in FLSM model well. Based on this, we further give an approximately analytic evaluation of $Q_{\mathrm{max}}$ by using the mapping method. We find the maximal charge is approximately $Q_{\mathrm{max}} \propto \tilde{g}^{-2}$ where $\tilde{g}$ is the gauge coupling of the dark $U(1)$ symmetry. The constraint on the value of gauge coupling $\tilde{g}$ is given by Eq.~\eqref{gmax} if the gauged Q-balls are produced by a FOPT. We discuss the relic density of gauged Q-ball DM formed during an electroweak FOPT. Even in the minimal electroweak FOPT model (SM plus singlet), the gauged Q-balls can comprise all the observed DM. And we have found that in order to satisfy the relic abundance of DM, the original DM asymmetry surprisingly coincides with the observed large lepton number asymmetry. The average charge and mass of gauged Q-ball DM can be varied by modifying the phase transition dynamics or the primordial DM asymmetry.
Besides, we give combined constraints on the gauged Q-ball DM from DM direct detection (Mica, XENON1T, Ohya), and astronomical observations (CMB, neutron stars, white dwarfs).
The formation process of gauged Q-ball DM during a FOPT also produces phase transition GW signals which could be detected by future GW experiments such as LISA, TianQin, and Taiji. The phase transition dynamics in the early Universe provide new formation mechanisms of various soliton DM or dynamical DM. For example, it is reasonable to discuss other species of soliton DM formed during FOPT, such as gauged Fermi-ball DM. The configuration of the Fermi-ball is different from the Q-ball because of the extra Fermi-gas degeneracy pressure. We leave these in our future works.

\paragraph{Acknowledgements:} 
The authors thank
Emin Nugaev, Julian Heeck, Michael Baker, Kiyoharu Kawana, Mikhail Smolyakov, Mikheil Sokhashvili, Yakov M. Shnir, Dmitry Levkov, Muhammad Fakhri Afif, Chris Verhaaren, Rebecca Riley, Sida Lu, Bingrong Yu, Jiahang Hu for helpful correspondence. And we thank the anonymous referee for the valuable advice.  This work was supported by the National Natural Science Foundation of China (NNSFC)
under Grant No. 12205387, by Guangdong Major Project of Basic and Applied Basic
Research (Grant No. 2019B030302001), and 
by KIAS Individual Grants under Grants No. PG021403 (PK). 

\bibliographystyle{JHEP}
\bibliography{gaugedqball}


\end{document}